\documentclass[article,shortnames,nojss]{jss}\usepackage[]{graphicx}\usepackage[]{color}
\makeatletter
\def\maxwidth{ %
  \ifdim\Gin@nat@width>\linewidth
    \linewidth
  \else
    \Gin@nat@width
  \fi
}
\makeatother

\usepackage{amsfonts}
\usepackage{amsmath}
\usepackage{bm}
\usepackage{bbm}
\newcommand{\ts}{^{\sf T}} 

\usepackage{thumbpdf,lmodern}

\usepackage{framed}

\author{Matteo Fasiolo \\University of Bristol 
   \And Simon N. Wood\\University of Bristol
   \AND Margaux Zaffran\\ENSTA Paris
   \And Rapha\"el Nedellec \\{\'E}lectricit{\'e} de France R\&D
   \And Yannig Goude \\{\'E}lectricit{\'e} de France R\&D}
\Plainauthor{Matteo Fasiolo, Second Author}

\title{\pkg{qgam}: Bayesian non-parametric quantile regression modelling in \proglang{R}}
\Plaintitle{qgam: Bayesian non-parametric quantile regression modelling in R}
\Shorttitle{Non-parametric regression modelling with \pkg{qgam}}

\Abstract{
Generalized additive models (GAMs) are flexible non-linear regression models, which can be fitted efficiently using the approximate Bayesian methods provided by the \pkg{mgcv} \proglang{R} package. While the GAM methods provided by \pkg{mgcv} are based on the assumption that the response distribution is modelled parametrically, here we discuss more flexible methods that do not entail any parametric assumption. In particular, this article introduces the \pkg{qgam} package, which is an extension of \pkg{mgcv} providing fast calibrated Bayesian methods for fitting quantile GAMs (QGAMs) in \proglang{R}. QGAMs are based on a smooth version of the pinball loss of \cite{koenker2005quantile}, rather than on a likelihood function, hence jointly achieving satisfactory accuracy of the quantile point estimates and coverage of the corresponding credible intervals requires adopting the specialized Bayesian fitting framework of \cite{fasiolo2017fast}. Here we detail how this framework is implemented in \pkg{qgam} and we provide examples illustrating how the package should be used in practice. 
}

\Keywords{Bayesian quantile regression, generalized additive models, regression splines, calibrated Bayes, fast Bayesian inference, \proglang{R}}
\Plainkeywords{Bayesian quantile regression, generalized additive models, regression splines, calibrated Bayes, fast Bayesian inference, R}

\Address{
  Matteo Fasiolo\\
  School of Mathematics\\
  University of Bristol\\
  University Walk\\
  BS8 1TW Bristol, United Kingdom\\
  E-mail: \email{matteo.fasiolo@gmail.com}\\
  URL: \url{http://www.bristol.ac.uk/maths/people/matteo-fasiolo/overview.html}
}
\IfFileExists{upquote.sty}{\usepackage{upquote}}{}
\begin{document}



\section[Introduction: additive quantile modelling in R]{Introduction: additive quantile modelling in \proglang{R}} \label{sec:intro}

Generalized additive models \citep[GAMs, ][]{hastie1990generalized} are flexible regression models, where the relation between the response distribution and several covariates is modelled non-parametrically, typically via spline bases expansions. In standard GAMs only one parameter of the response distribution (typically the location) is modelled additively, but \cite{rigby2005generalized} developed methods for handling more flexible generalized additive models for location, scale and shape (GAMLSS), where potentially all the parameters of the response distribution are allowed to depend on the covariates. The quantile GAM models (QGAMs) described in this work provide even more flexibility by modelling the quantiles of conditional response distribution individually, thus avoiding any parametric assumption on the distribution of the response variable.

The purpose of this article is to discuss methods and software for QGAM modelling in \proglang{R}. In particular, we focus on the \pkg{qgam} \proglang{R} package \citep{qgampackage}, which implements the fast calibrated Bayesian fitting methods proposed by \cite{fasiolo2017fast}. The \pkg{qgam} package is an extension of the recommended \pkg{mgcv} package \citep{mgcvpackage}, which provides tools for building, fitting and visualizing GAM and GAMLSS models. A new package is required, because QGAMs can not be handled using the standard GAM methods implemented by \pkg{mgcv}. In particular, QGAMs are based on the pinball loss function \citep{koenker1978regression}, rather than on a probabilistic model of the response distribution, and the absence of a likelihood function impedes direct application of the Bayes' rule when updating the prior distribution on the regression coefficients given the observed responses. While this problem can be overcome by adopting the coherent Bayesian belief updating framework of \cite{bissiri2016general}, which effectively leads to the application of Bayes' rule using a loss-based pseudo-likelihood,  na\"ively plugging such pseudo-likelihood into standard Bayesian fitting methods can lead to inaccurate quantile estimates and inadequate coverage of the corresponding credible intervals as discussed, for instance, in \cite{yang2016posterior} and \cite{sriram2015sandwich}. To avoid such issues, \pkg{qgam} implements the calibrated Bayesian methods of \cite{fasiolo2017fast}, which explicitly aim at selecting the `learning rate' tuning parameter of the loss so as to achieve near-nominal frequentist coverage of the quantile credible intervals. Furthermore, \pkg{qgam} bases quantile regression on a smoothed version of the pinball loss, which enables the adoption of fast maximum a posterior (MAP) and empirical Bayes methods to estimate the regression coefficient and select their prior variance hyper-parameters, respectively. The smoothness of the new loss is determined by minimizing the asymptotic mean squared error (MSE) of the estimated regression coefficients, approximated using a location-scale GAM model.

To our best knowledge and at the time of writing, the QGAM model fitting framework proposed by \cite{fasiolo2017fast} is the only one able to estimate all the regression coefficients and prior hyper-parameters using fast direct optimization methods and to provide, at no extra computational cost, credible intervals which achieve adequate coverage for tail quantiles. It should be clarified that, when we describe \pkg{qgam} as ``fast'', we mean that it is the fastest additive quantile regression method we know of, which has the properties just described. Hence, for example, the \code{rqss} function provided by the \pkg{quantreg} package \citep{koenker2013quantreg} might be faster, if the smoothing parameters are known or if the model of interest contains at most one or two such parameters, which could be selected by minimizing a model selection criterion numerically, as discussed in \cite{koenker2011additive}, Section 3. Another alternative to the calibrated Bayesian estimation methods provided by \pkg{qgam} is additive quantile regression via gradient boosting, available in \proglang{R} via the \pkg{mboost} package \citep{mboostpackage}. \cite{fasiolo2017fast} shows that boosting can provide accurate point estimates, but selecting the optimal number of boosting steps requires running a computationally intensive cross-validation routine. Furthermore, the uncertainty of the estimated quantiles must be obtained by bootstrapping. \cite{waldmann2013bayesian} propose Bayesian methods for estimating QGAMs via the \pkg{BayesX} stand-alone software \citep{brezger2003bayesx}, with a specific quantile regression family being accessible from \proglang{R} using the \pkg{bamlss} package \citep{bamlsspackage}. However, their proposal is based on Markov chain Monte Carlo (MCMC) methods, which are slower than the direct optimization methods adopted here, and the resulting credible intervals struggle to achieve nominal frequentist coverage for tail quantiles, as detailed in \cite{waldmann2013bayesian}. Another alternative is the \pkg{VGAM} \citep{yee2008vgam} \proglang{R} package but, as for \verb|quantreg|, the prior smoothing hyper-parameters have to be selected manually. \cite{lin2013variable} focus on variable selection, not smoothing, hence the corresponding software can not be considered an alternative to \pkg{qgam}. 

The rest of the paper is structured as follows. In Section \ref{sec:modMetSoft} we first present the basic structure of QGAM models, then we show how such models can be fitted using the fast calibrated Bayesian methods of \cite{fasiolo2017fast} and we explain how the fitting framework is implemented in the \pkg{qgam} package. In Section \ref{sec:examples} we illustrate how the package can be used for quantile GAM modelling. In particular, the first example introduces the basic features of the package, while the second example considers a more realistic application, focused on electricity demand forecasting using functional effects. In Section \ref{sec:conclusions} we outline some promising directions for future work on additive quantile modelling with \pkg{qgam}.



\section{Models, methods and software} \label{sec:modMetSoft}

\subsection{General structure of additive quantile regression models}

Let $y$ be a continuous random variable (r.v.) with conditional distribution $p(y|{\bm x})$, where ${\bm x}$ is a $p$-dimensional vector of covariates. Define also the conditional quantiles of $p(y|{\bm x})$ by $\mu_\tau(\bm x) = F^{-1}(\tau|{\bm x})$, where $\tau \in (0, 1)$ and $F^{-1}$ is the inverse conditional cumulative distribution function (c.d.f.) of $y$. In quantile regression \citep{koenker1978regression} the conditional quantiles are modelled individually, without specifying a model for $p(y|{\bm x})$. Direct quantile estimation is generally achieved by exploiting the following alternative definition of a quantile  
\begin{equation} \label{eq:quantDef}
\mu_\tau({\bm x}) = \underset{\mu}{\text{argmin}} \, \mathbb{E}\{ \rho_\tau(y - \mu) | \bm x \},
\end{equation}
where 
$$
\rho_\tau(z) = (\tau-1)\frac{z}{\sigma} \mathbbm{1}(z < 0) + \tau \frac{z}{\sigma} \mathbbm{1}(z \geq 0),
$$
is the scaled version of the so-called `pinball' or `check' loss. $\sigma > 0$ is a scale parameter, which can potentially depend on the covariates $\bm x$. Its role will be clarified in Section \ref{sec:BayesianUpdate}. In \pkg{qgam} the pinball loss, which is piecewise linear and has discontinuous derivatives, is replaced by the extended log-f (ELF) loss \citep{fasiolo2017fast}, that is
\begin{equation} \label{eq:ELFloss}
\tilde{\rho}_\tau(z) = (\tau - 1)\frac{z}{\sigma} + \lambda \log(1+e^{\frac{z}{\lambda\sigma}}), 
\end{equation}
where $\lambda > 0$, and the pinball loss is recovered as $\lambda \rightarrow 0^+$. The ELF loss is a smoothed version of the pinball loss which, as will be explained later, has the advantages of leading to more accurate quantile estimates (if $\lambda$ is tuned adequately) and of enabling the use of efficient computational methods for model fitting.

In this work we assume that the conditional quantiles have an additive structure, that is 
$$
\mu({\bm x}) = \sum_{j = 1}^J f_j({\bm x}),
$$
where the $f_j$'s are parametric, random or smooth effects. The latter are built using spline bases expansions, so the $j$-th effect can be written
$$
f_j({\bm x}) = \sum_{k=1}^{K_j} b_k^j(\bm x)\beta_k^j,
$$
where $b_1^j, \dots, b_{K_j}^j$ are the spline basis function used to built the $j$-th effect and $\beta_1^j, \dots, \beta_k^j$ are the corresponding regression coefficients. The basis functions are known and fixed, while the regression coefficients must be estimated. Note that the quantile $\mu$ depends both on $\bm x$ and $\bm \beta$, but in the following we will refer to it using $\mu({\bm x})$ or $\mu({\bm \beta})$, depending on the context.

In a Bayesian framework, the complexity of the smooth and random effects is controlled using a prior distribution on the regression coefficients, which we indicate with $p({\bm \beta})$. In this work we assume that the prior is an improper multivariate Gaussian distribution, centered at $\bm 0$ and with positive semi-definite precision matrix ${\bf S}^{\bm \gamma} = \sum_{l=1}^m \gamma_l {\bf S}_l$. Here the ${\bf S}_l$'s are positive semi-definite matrices, scaled by the positive parameters $\bm \gamma = \{\gamma_1, \dots, \gamma_l\}$. We assume that the ${\bf S}_j$'s are given, while the vector $\bm \gamma$ needs to be selected. The ${\bf S}_j$'s related to random effects are often simple diagonal matrices, while the ${\bf S}_j$'s related to smooth effects have more complex structures, aimed at penalizing departures from (some definition of) smoothness. Hence, increasing $\bm \gamma$ leads to a prior where the random effects coefficients are more concentrated around zero and the smooth effects less wiggly. In the following we refer to $\bm \gamma$ as the vector of smoothing parameters, with the understanding that some of its elements could instead be controlling the prior precision of the random effects. In general, there needs not to be a one-to-one correspondence between the smoothing parameters and the effects because, for example, a smoothing parameter might be determining the prior precision of several effects or the complexity of a single effect might be controlled using multiple smoothing parameters. See \cite{wood2006generalized} for a detailed account on smooth and random effects model structures, in an additive modelling context.   

This section outlined the basic QGAM modelling framework, the next two will detail how such models can be estimated within a fast calibrated Bayesian framework.

\subsection{Performing the Bayesian update under the ELF loss} \label{sec:BayesianUpdate}

Let us indicate with $p(y|{\bm \beta})$ the true conditional distribution of the response, where the dependency on $\bm x$ has been temporarily suppressed to simplify the notation, and indicate with $p({\bm \beta})$ the prior distribution, which is implicitly a function of the smoothing parameters $\bm \gamma$. Assume for the moment that the latter have been fixed to some value, so that the prior is given. Recall that we are basing additive quantile regression on the ELF loss, rather than on a probabilistic model for $p(y|{\bm \beta})$. This is an impediment to performing Bayesian inference for QGAMs, as the lack of a likelihood function prevents us from applying Bayes' rule to update $p({\bm \beta})$ to the corresponding posterior, $p({\bm \beta}|y)$. We address the issue by adopting the belief updating (BU) framework of \cite{bissiri2016general}, which allows to perform the Bayesian update using a general loss function, rather than a likelihood. In particular, as detailed in \cite{fasiolo2017fast}, applying the BU framework to the ELF loss leads to the following update formula
\begin{equation} \label{eq:GibbsPost}
p(\bm \beta|y) \propto \tilde{p}_\tau\{y - \mu(\bm \beta)\} p({\bm \beta}),
\end{equation}
where 
\begin{equation} \label{eq:logFLambda}
\tilde{p}_\tau(y - \mu) = \frac{e^{-\tilde{\rho}_\tau\{y - \mu\}}}{\int e^{-\tilde{\rho}_\tau\{y - \mu\}} dy} = \frac{e^{(1-\tau)\frac{y-\mu}{\sigma}}(1+e^{\frac{y-\mu}{\lambda\sigma}})^{-\lambda}}{\lambda\sigma\text{Beta}\big[\lambda(1-\tau),\lambda\tau\big]},
\end{equation}
is the probability density function (p.d.f.) of the ELF distribution, which is an extension of the log-f distribution of \cite{jones2008class}. Here $\lambda$ determines the smoothness of the loss, while $1/\sigma$ plays the role of a `learning rate', determining the relative weights of the loss-based pseudo-likelihood and the prior. In fact, letting $\lambda \rightarrow 0^+$ leads to $p(\bm \beta|y) \propto \exp\{-\rho_\tau\{y - \mu\} / \sigma\} p(\bm \beta)$ and, as $1/\sigma$ increases, the loss-based likelihood progressively dominates the prior, thus leading to faster learning and a higher risk of overfitting. Decreasing $1/\sigma$ has the opposite effect. In contexts where the variance of $y$ varies strongly with $\bm x$, it can be advantageous to let $\sigma$ depend on $\bm x$ via the decomposition $\sigma(\bm x) = \sigma_0 \tilde{\sigma}(\bm x)$, where $\sigma_0$ is the baseline or average learning rate and $\tilde{\sigma}(\bm x)$ is its $\bm x$-dependent component.

Having defined (\ref{eq:GibbsPost}) which, following \cite{syring2015scaling}, we refer to as the Gibbs posterior, the next section outlines how the fast calibrated Bayesian methods proposed by \cite{fasiolo2017fast}, and implemented by \pkg{qgam}, can be used to fit ELF-based QGAMs.

\subsection{Fast calibrated Bayesian model fitting methods} \label{sec:modelFitting}

We fit quantile GAMs using three nested optimization routines, in particular:
\begin{enumerate}
\item in the outer iteration, the baseline learning rate $1/\sigma_0$ is selected by minimizing a calibration loss function numerically;
\item for fixed $\sigma_0$, the loss smoothness $\lambda$ and the $\bm x$-dependent component of the learning-rate $\tilde{\sigma}(\bm x)$ are determined using closed-form expressions, while the smoothing parameters $\bm \gamma$ are selected by numerically optimizing an intermediate criterion;
\item for fixed $\sigma(\bm x) = \sigma_0\tilde{\sigma}(\bm x)$, $\lambda$ and $\bm \gamma$, the regression coefficients $\bm \beta$ are estimated by numerically optimizing an inner criterion.
\end{enumerate}
Of course parameter $\tau \in (0, 1)$, which indicates the quantile of interest, is kept fixed throughout. The fact that the three iterations are nested, not sequential, means that evaluating the outer objective function requires solving the intermediate optimization problem and, in turn, each step of the latter entails running the inner optimization to convergence. In the following we provide a methodological outline of the three iterations, going from the inner to the outer one, and in Section \ref{sec:softImpl} we describe how the whole procedure is implemented in \pkg{qgam}.

\subsubsection{Inner iteration: MAP estimation of the regression coefficients}
 
The inner iteration estimates the regression coefficients using efficient maximum a posteriori (MAP) methods, for fixed $\bm \gamma$, $\sigma(\bm x)$ and $\lambda$. In particular, let ${\bf y} = \{y_1, \dots, y_n\}$ be the vector of observed responses and note that the conditional quantile is modelled by $\mu({\bm x}_i) = {\bf x}_{i}\ts{\bm \beta}$, where ${\bf x}_i$ is the $i$-th row of the $n \times d$ design matrix $\bf X$, containing the spline basis functions evaluated at $\bm x_i$. It is easy to show that maximizing the logarithm of the ELF-based Gibbs posterior (\ref{eq:GibbsPost}) is equivalent to minimizing the criterion
\begin{equation} \label{eq:DevCrit}
\tilde{V}_D\{\bm \beta, \bm \gamma, \sigma(\bm x), \lambda\} = \sum_{i=1}^n \text{Dev}_i\left\{\bm \beta, \sigma({\bm x}_i), \lambda \right\} + \sum_{l=1}^m \gamma_j \bm \beta\ts {\bf S}_j \bm \beta,
\end{equation}
%
where $\text{Dev}_i\left\{\bm \beta, \sigma({\bm x}_i), \lambda \right\}$ is the $i$-th deviance component, based on the ELF density (\ref{eq:logFLambda}). Criterion (\ref{eq:DevCrit}) is smooth and can be minimized efficiently with respect to (w.r.t.) $\bm \beta$ using a penalized iteratively re-weighted least squares (PIRLS) algorithm, as detailed in \cite{fasiolo2017fast}.

\subsubsection{Intermediate iteration: selection of the smoothing parameters and ELF loss smoothness}

The intermediate iteration selects the smoothing parameters by maximizing a Laplace approximation to the ELF-based marginal likelihood. The latter is
\begin{equation}\label{eq:ELFmargLik}
\log p\{{\bf y} | {\bm \gamma}, \sigma(\bm x), \lambda\} = \int \tilde{p}_\tau(y_i - {\bf x}_{i}\ts{\bm \beta}) p({\bm \beta}|{\bm \gamma}) d \bm \beta,
\end{equation}
where we have made explicit the dependence of the prior on the smoothing parameters. Applying a Laplace approximation to the intractable integral (\ref{eq:ELFmargLik}) leads to the following Laplace approximate marginal likelihood (LAML) criterion
\begin{align}\label{eq:LAML}
G\{{\bm \gamma}, \sigma(\bm x), \lambda\} = \,  - & \frac{1}{2}\tilde{V}_D\{\hat{\bm \beta}, \bm \gamma, \sigma(\bm x), \lambda\} + \tilde{\text{ll}}\{\sigma(\bm x), \lambda\} \nonumber  \\  - &  \frac{1}{2} \Big [ \log |{\bf X\ts  W  X +  S^{\bm \gamma}| - \log|S^{\bm \gamma}|_+} \Big] + \frac{M_p}{2} \log(2 \pi),
\end{align}
where $\hat{\bm \beta}$ is the minimizer of (\ref{eq:DevCrit}), estimated using the inner iteration, $\tilde{\text{ll}}\{\sigma(\bm x), \lambda\}$ is the saturated log-likelihood corresponding to the ELF density (\ref{eq:logFLambda}), $M_p$ is the dimension of the null-space of ${\bf S}^{\bm \gamma}$ and $|{\bf S}^{\bm \gamma}|_+$ is the product of its non-zero eigenvalues. \cite{fasiolo2017fast} focus on a marginal loss criterion, which is equal to the negative of (\ref{eq:ELFmargLik}) up to a normalization constant, to stress that $\sigma_0$ should not be selected by jointly maximizing (\ref{eq:LAML}) w.r.t. $\bm \gamma$ and $\sigma_0$. This would be appropriate if $\sigma_0$ were the scale parameter of a standard GAMs, but that the outer calibration procedure for $\sigma_0$ should be used instead. Here we prefer to adopt a likelihood-based terminology to emphasize that (\ref{eq:LAML}) can be maximized w.r.t. $\bm \gamma$ using the numerically stable methods of \cite{wood2016smoothing}, which are aimed at standard probabilistic GAMs. 

Before describing the outer iteration, we need to specify how $\tilde{\sigma}({\bm x})$ and $\lambda$, which are held constant throughout the intermediate and inner iterations, are determined for fixed $\sigma_0$. Recall that $\sigma(\bm x) = \sigma_0\tilde{\sigma}(\bm x)$ and define the scaled loss bandwidth $h(\bm x) = \sigma(\bm x)\lambda$. Assume that the responses follow the location-scale model $y_i|{\bm x_i} \sim \alpha({\bm x}_i) + \kappa({\bm x}_i) z_i$, where the $z_i$'s are i.i.d with $\mathbb{E}(z|{\bm x})=0$ and $\text{var}(z|{\bm x})=1$. \cite{fasiolo2017fast} shows that, under the above location-scale model and further assumptions specified therein, the asymptotic MSE of the regression coefficients is minimized by
\begin{equation} \label{eq:optBanZ}
\tilde{h}^*(\bm x) = \left[\frac{d}{n}\frac{9f_z\{F_z^{-1}(\tau)\}}{\pi^4f'_z\{F_z^{-1}(\tau)\}^2}\right]^{\frac{1}{3}}\kappa(\bm x),
\end{equation}
where $f_z$, $f'_z$ and $F_z$ are the p.d.f. of $z$, its first derivative and its c.d.f., $F_z^{-1}(\tau)$ is the $\tau$-th quantile of $z$ and $d$ the dimension of $\bm \beta$. Having determined $\tilde{h}^*(\bm x)$, we impose $n^{-1} \sum_i \tilde{\sigma}({\bm x}_i) = 1$ so that $\lambda = n^{-1} \sum_i \tilde{h}^*({\bm x}_i) / \sigma_0$. The idea is to use $\tilde{\sigma}({\bm x})$ to modulate the baseline scaled loss smoothness, $\lambda\sigma_0$, and learning rate, $1/\sigma_0$, to make them respectively directly and inversely proportional to the conditional variance of $y$. Hence, for fixed $\sigma_0$, we obtain $\tilde{\sigma}({\bm x})$ and $\lambda$ using the formulas just mentioned. This, of course, requires estimates of $\alpha(\bm x)$ and $\kappa(\bm x)$ in the location-scale model, as well as of $f_z$ and $F_z$. Such estimates need to be obtained only once, before initiating the nested QGAM fitting procedure described here, hence we will discuss the specific approach we follow in Section \ref{sec:softImpl}, where we also describe its software implementation.

\subsubsection{Outer iteration: selection of the learning rate by calibrated Bayes}

Recall that $1/\sigma_0$ is the baseline learning rate, which determines the relative weight of the loss and the prior in the Gibbs posterior (\ref{eq:GibbsPost}). Increasing $1/\sigma_0$ leads to faster learning, that is wigglier fitted quantile curves and narrower posterior credible intervals. \cite{fasiolo2017fast} selects $\sigma_0$ using a calibration procedure aimed at guaranteeing that the credible intervals for $\mu(\bm x)$ approximately achieve the correct frequentist coverage. This is achieved by minimizing  
\begin{equation} \label{eq:IKLestim}
\hat{\text{IKL}}(\sigma_0) = n^{-1}\sum_{i=1}^n \bigg[ \frac{\hat{{v}}_s({\bm x}_i)}{v({\bm x}_i)} + \log\frac{v({\bm x}_i)}{\hat{{v}}_s({\bm x}_i)} \bigg]^\zeta, 
\end{equation}
which is an estimate of the integrated Kullback-Leibler (IKL) divergence 
\[
\text{IKL}(\sigma_{0})=\int\text{KL}\big[\text{N}\{\mu({\bm x}),{v}_s({\bm x})\}, \text{N}\{\mu({\bm x}),v({\bm x})\}\big]^{\zeta}p({\bm x})d{\bm x}\propto\int\bigg\{\frac{{v}_s({\bm x})}{v({\bm x})}+\log\frac{v({\bm x})}{{v}_s({\bm x})}\bigg\}^{\zeta}p({\bm x})d{\bm x}.
\]
Here $\text{KL}(\cdot, \cdot)$ and $\text{N}(\cdot, \cdot)$ indicate respectively the KL divergence and the univariate Gaussian distribution, $\zeta$ is a positive parameter which we fix to $1/2$, $v({\bm x})={\bf x}\ts{{\bf V}}{\bf x}$ and ${v}_s({\bm x})={\bf x}\ts{{\bf V}}_s{\bf x}$ are the posterior variance of $\mu({\bm x})$ under two alternative posterior covariance matrices for $\bm \beta$. In particular ${{\bf V}}=({\cal I}+{\bf S}^{\bm{\gamma}})^{-1}$ and ${{\bf V}}_s=({\cal I}\bm{\Sigma}_{\nabla}^{-1}{\cal I}+{\bf S}^{\bm{\gamma}})^{-1}$, where ${\cal I}$ is the negative Hessian of the ELF-based log-likelihood and $\bm{\Sigma}_{\nabla}=\text{cov}[\nabla_{\bm \beta}\tilde{\rho}\{y - \mu(\bm \beta)\}|_{\bm \beta = \hat{\bm \beta}}]$ is the covariance matrix of the gradient of the ELF loss, under the data generating process. $\hat{{v}}_s({\bm x}_i)$ is an estimate of ${v}_s({\bm x}_i)$, based on the regularized estimator of ${{\bf V}}_s$ described in \cite{fasiolo2017fast} (who use a different notation and indicate ${v}_s({\bm x}_i)$ and ${{\bf V}}_s$ by $\tilde{v}({\bm x}_i)$ and $\tilde{{\bf V}}$, respectively). The IKL objective function is generally smooth and convex, hence we minimize it efficiently using Brent's method \citep{brent2013algorithms}. 

\cite{fasiolo2017fast} propose also an alternative calibration procedure, based on another version of the IKL loss where $\hat{v}_s({\bm x})$ is substituted by the sample pointwise variance of the estimated quantile $\hat{\mu}({\bm x})$, estimated by bootstrapping. Such a procedure requires re-fitting the model to several bootstrap samples, hence it is computationally more expensive than minimization of the IKL loss defined above, but it can lead to better coverage in small samples. We refer to \cite{fasiolo2017fast} for explanations regarding how minimizing either version of the IKL loss w.r.t. $\sigma_0$ leads to better frequentist coverage of the posterior credible intervals, relative to joint maximization of LAML (\ref{eq:LAML}) w.r.t. $\sigma_0$ and $\bm \gamma$.  

This section outlined the Bayesian QGAM fitting framework of \cite{fasiolo2017fast} from a methodological point of view. The next section
explains how the fitting framework just described is implemented in the \pkg{qgam} package.

\subsection{Software implementation of the fitting framework} \label{sec:softImpl}

The \pkg{qgam} package is an extension of \pkg{mgcv} providing additional tools for handling QGAMs. Here we focus mainly on how the package implements the fitting framework of \cite{fasiolo2017fast}, while in Section \ref{sec:examples} we provide usage examples. 

\begin{table}[h]

\centering 
\begin{tabular}{l l} 
\hline 
Function name&\multicolumn{1}{l}{Description} \\ 
\hline 
\code{qgam} & Fits a QGAM model for a single quantile $\tau$. It is analogous to the \code{gam} \\
            & function in \pkg{mgcv}, hence any smooth effect type available in \code{gam} is also \\
            & available in \code{qgam}. \\ 
\code{mqgam} & Fits the same QGAM model to a vector of $k$ quantiles $\tau_1, \dots, \tau_k$ more \\
             & efficiently than by calling \code{qgam} repeatedly. \\
\code{tuneLearnFast} & Selects the learning rate by minimizing (\ref{eq:IKLestim}) using Brent's method. \\
\code{tuneLearn} & Evaluates (\ref{eq:IKLestim}) on a grid of values of $\sigma_0$.  \\
\code{check} & Given a model fitted using \code{qgam}, produces some diagnostics plots. \\
\code{qdo} & Is a wrapper for using standard generics (e.g., \code{summary}) on the output \\
           & of \code{mqgam}. Needed because \code{mqgam} does not output an object of class \code{gam}. \\
\hline 
\end{tabular}
\label{tab:hresult}
\caption{Main functions provided by \pkg{qgam}.} 
\label{tab:mytable}
\end{table}

Table \ref{tab:mytable} lists the main functions provided by \pkg{qgam}. The most commonly used one is itself called \code{qgam} and has the following arguments
\begin{Code}
qgam(form, data, qu, lsig = NULL, err = NULL,
  multicore = !is.null(cluster), cluster = NULL, ncores = detectCores() - 1, 
  paropts = list(), control = list(), argGam = NULL)
\end{Code}
Arguments \code{form} and \code{data} have the same meaning as in \code{mgcv::gam} (henceforth just \code{gam}), \code{qu} is the quantile of interest $\tau$, \code{lsig} is $\log{\sigma_0}$ and \code{err} is a positive parameter allowing to set the ELF loss smoothness manually. By default \code{qgam} determines the loss smoothness automatically, using the methods described in Section \ref{sec:modelFitting}, hence most users will not need to use argument \code{err}. However, Appendix \ref{sec:lossBand} provides some guidelines for users who decide to select the loss smoothness manually. In \code{qgam} version \code{1.3.2}, arguments \code{multicore}, \code{cluster}, \code{ncores} and \code{paropts} are relevant only when the learning rate is calibrated by bootstrapping, and can be used to perform bootstrap re-fitting in parallel. \code{control} is a list of control parameters useful, for instance, to choose the type of calibration used and to suppress the text output printed by \verb|qgam|, while \code{argGam} is a list of arguments to be passed to \code{gam}.

To describe how \pkg{qgam} implements the fitting framework of \cite{fasiolo2017fast}, assume that we want to fit a QGAM for quantile $\tau = 0.5$, with model formula \code{myForm = list(y ~ s(x1) + s(x2), ~ s(x3))} and using data from a \code{data.frame} called \code{myDat}. Here \code{y ~ s(x1) + s(x2)} is the model used for the median, that is $\mu_\tau(\bm x) = f_1(x_1) + f_2(x_2)$, where $f_1$ and $f_2$ are two smooth effects constructed, by default, using ten thin plate splines basis function. The second element of the formula, \code{~ s(x3)}, is the model used for the variance $\kappa(\bm x)$ in the preliminary location-scale GAM model fit. The first thing to point out is that, if $\sigma_0$ and the bandwidth $h(\bm x)$ were known, QGAM models could be fitted using \code{gam} directly. In fact, we could fit a median quantile model using 
\begin{Code}
gam(myForm[[1]], data = myDat, 
  family = elf(theta = 1, qu = 0.5, co = rep(1, nrow(myDat))))
\end{Code}
where \code{qgam::elf} is a family providing the ELF log-likelihood function and its derivatives. Its argument \code{theta} and \code{co} correspond respectively to $\log\sigma_0$ and $h(\bm x)$, which have been arbitrarily fixed to 1 here. Hence, given $\sigma_0$ and $h(\bm x)$, the intermediate and inner iterations for selecting $\bm \gamma$ and estimating $\bm \beta$ take place inside \code{gam}. But these parameters are generally unknown, so \code{qgam} determines $\sigma_0$ using the outer iteration and $h(\bm x)$ via the preliminary location-scale GAM fit. In particular, upon calling 
\begin{Code}
qgam(myForm, data = myDat, qu = 0.5)
\end{Code}                  
the \code{qgam} function executes the following pseudo-code: 
\begin{enumerate}
\item estimate $\alpha(\bm x)$ and $\kappa(\bm x)$ by fitting a Gaussian location-scale GAM
\begin{Code}
gam(myForm, data = myDat, family = gaulss)
\end{Code}
where \code{mgcv::gaulss} is a Gaussian location-scale family.
\item numerically fit the sinh-arch density of \cite{jones2009sinh} to the standardized Gaussian GAM residuals, to provide estimates of $f_z$, $f'_z$ and $F_z^{-1}(\tau)$.
\item estimate $\tilde{h}^*(\bm x)$ by plugging the estimates obtained in previous two steps into (\ref{eq:optBanZ}), where
      $d$ is set to the number of effective degrees of freedom used to model $\alpha(\bm x)$ in step 1.
\item call the \code{tuneLearnFast} function, which minimizes the IKL loss w.r.t. $\log \sigma_0$ using an outer Brent algorithm. 
      For each trial value of $\log \sigma_0$: 
      \begin{enumerate}
      \item call \code{gam} with arguments \code{formula = myForm[[1]]} and \code{family = elf}, the latter having 
            parameter \code{theta} fixed to the current value of $\log\sigma_0$ and \code{co} 
            fixed to $\tilde{h}^*(\bm x)$. ELF parameters $\lambda$ and $\sigma(\bm x)$ are obtained within \code{elf} by 
            $\lambda = n^{-1} \sum_i \tilde{h}^*({\bm x}_i) / \sigma_0$ and then $\sigma(\bm x_i) = \tilde{h}^*({\bm x}_i) / \lambda$.
            The \code{gam} call performs the intermediate iteration for selecting $\bm \gamma$ and the inner 
            iteration for estimating $\bm \beta$.
      \item Estimate the IKL loss using (\ref{eq:IKLestim}), where $v(x)$ is easily obtained from the \code{gam} 
            output, while ${v}_s(\bm x)$ is estimated using the regularized estimator of \cite{fasiolo2017fast}. 
      \end{enumerate}
      Upon convergence, return the value of $\log \sigma_0$ minimizing the IKL loss.
\item Call \code{gam} as in step 4(a), with \code{elf} parameter \code{theta} fixed to the value of $\log \sigma_0$ returned by \code{tuneLearnFast}, and \code{co} set to $\tilde{h}^*(\bm x)$.
\item Return the output of the last \code{gam} call, with the output of \code{tuneLearnFast} call stored in its \code{$calibr} slot.
\end{enumerate}

Note that a variety of different density estimators could be used in step 2 above. We use the sinh-arch distribution because it provides a good balance between flexibility and stability. However, future versions of \code{qgam} might allow user to provide their own density estimator. Similar considerations hold for the Gaussian GAM used in step 1. The \code{tuneLearnFast} function of step 4 has exactly the same arguments as \code{qgam}, and it returns a list containing diagnostic information regarding the outer iteration and the value of $\sigma_0$ which minimizes the IKL loss. The output of \code{qgam} is an object of class \code{c("qgam", "gam", "glm", "lm")}, which can be manipulated using any of the \code{S3} methods available for objects of class \code{"gam"}. Version 1.3.2 of the \pkg{qgam} package does not provide any \code{S4} method or class.

\section{Examples} \label{sec:examples} 

\subsection{A basic example: the motorcycle data set} \label{sec:motorExample}  

Here we consider the classic motorcycle accident data set of \cite{silverman1985some}, available in the MASS package \citep{ripley2013package}. We start by loading the data and fitting a QGAM for quantile $\tau = 0.9$ as follows

\begin{figure}[t]
\centering
\begin{Schunk}

\includegraphics[width=\maxwidth]{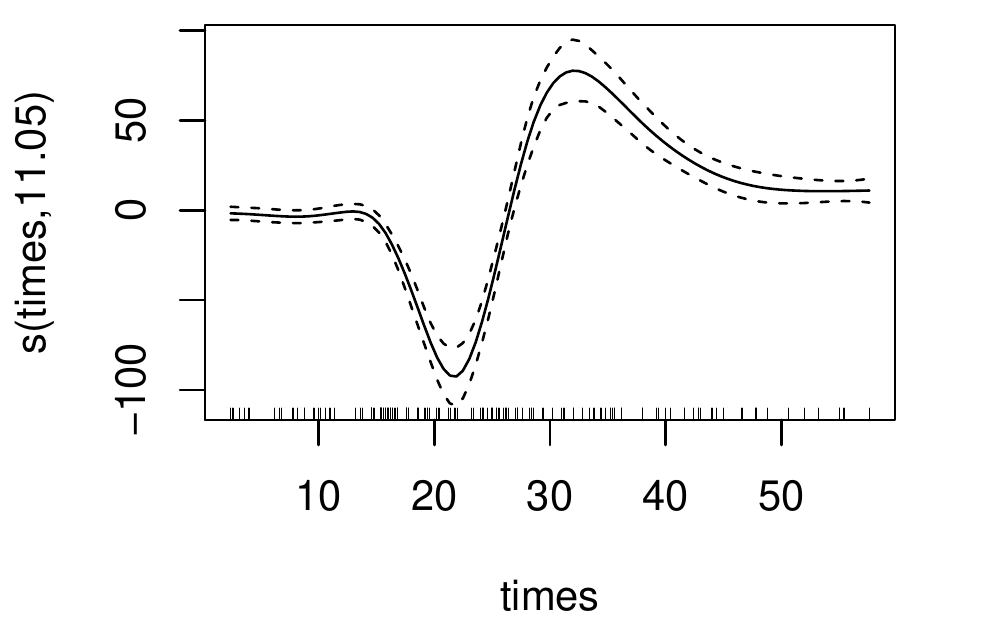} \end{Schunk}
\caption{\label{fig:mcycleEff1} Fitted smooth effect of time on quantile 0.9 of the \code{mcycle} data set, with 95\% credible intervals obtained using a Gaussian approximation to the posterior.}
\end{figure}

\begin{Schunk}
\begin{Sinput}
R> library("MASS")
R> library("qgam")
R> fitCycle1 <- qgam(list(form = accel ~ s(times, k = 20, bs = "ad"), 
+    ~ s(times)), data = mcycle, qu = 0.9)
\end{Sinput}
\end{Schunk}
We are using the model $\mu_\tau(\text{time}) = f(\text{time})$, where $f(\cdot)$ is an adaptive smooth effect built using 20 P-splines basis functions \citep{eilers1996flexible}. We use an adaptive smooth because, as shown in Figure \ref{fig:mcycleEff1}, the shape of the conditional quantile varies sharply between roughly 10 and 40 milliseconds, but it is otherwise quite flat. Recall that the second part of the formula, \code{~ s(times)}, is not used to model the quantile, but the variance $\kappa(\bm x)$ of the response within the location-scale GAM model.

Given that \code{fitCycle1} inherits from class \code{"gam"}, generic function calls such as
\begin{Schunk}
\begin{Sinput}
R> predict(fitCycle1)[1:5] 
\end{Sinput}
\begin{Soutput}
          1           2           3           4           5 
0.460372275 0.421737850 0.271621363 0.146774904 0.005606131 
\end{Soutput}
\end{Schunk}
dispatch to the relevant method (here \code{predict.gam}). Similarly, the fitted effects in the quantile model can be plotted using
\begin{Schunk}
\begin{Sinput}
R> plot(fitCycle1)
\end{Sinput}
\end{Schunk}
which dispatches to \code{plot.gam}, and produces the plot shown in Figure \ref{fig:mcycleEff1}. 
To save memory, no component of the Gaussian location-scale fit is stored in the output of \code{qgam}, but the selected values of the parameters $\log \sigma_0$ and $h(\bm x)$ can be extracted by
\begin{Schunk}
\begin{Sinput}
R> fitCycle1$family$getTheta()
\end{Sinput}
\begin{Soutput}
     0.9 
1.237221 
\end{Soutput}
\begin{Sinput}
R> fitCycle1$family$getCo()[1:5]
\end{Sinput}
\begin{Soutput}
[1] 0.2791352 0.2728345 0.2549963 0.2442300 0.2346063
\end{Soutput}
\end{Schunk}

Above we fitted a single quantile, but multiple quantiles can be estimated at once using the following function call
\begin{Schunk}
\begin{Sinput}
R> fitCycleM <- mqgam(list(form = accel ~ s(times, k = 20, bs = "ad"), 
+    ~ s(times)), data = mcycle, qu = c(0.1, 0.25, 0.5, 0.75, 0.9))
\end{Sinput}
\end{Schunk}
where the \code{mqgam} function fits a QGAM to each of the five quantiles. The main advantage of using \code{mqgam}, rather than several calls to \code{qgam}, is that it saves memory by storing a single version of potentially large objects, such as the model matrix, for all the fitted QGAMs. The five fitted QGAM models can be found in the list \code{fitCycleM$fit}, but calling functions such as \code{plot(fitCycleM$fit[[1]])} would trigger an error because some object elements are missing. The output of \code{mqgam} should instead be manipulated using the \code{qdo} function, for example
\begin{Schunk}
\begin{Sinput}
R> qdo(fitCycleM, qu = 0.25, fun = summary)
\end{Sinput}
\begin{Soutput}

Family: elf 
Link function: identity 

Formula:
accel ~ s(times, k = 20, bs = "ad")

Parametric coefficients:
            Estimate Std. Error z value Pr(>|z|)    
(Intercept)  -42.739      2.016   -21.2   <2e-16 ***
---
Signif. codes:  0 '***' 0.001 '**' 0.01 '*' 0.05 '.' 0.1 ' ' 1

Approximate significance of smooth terms:
           edf Ref.df Chi.sq p-value    
s(times) 10.55  11.83  837.3  <2e-16 ***
---
Signif. codes:  0 '***' 0.001 '**' 0.01 '*' 0.05 '.' 0.1 ' ' 1

R-sq.(adj) =  0.735   Deviance explained = 89.2%
-REML = 570.51  Scale est. = 1         n = 133
\end{Soutput}
\end{Schunk}
shows the output of \code{summary.gam} for the QGAM fitted to quantile $\tau = 0.25$. Similarly, the fitted effect for $\tau = 0.1$ can be plotted by \code{qdo(fitCycleM, qu = 0.1, fun = plot)}. 

More convenient methods for plotting multiple QGAMs jointly are provided by the \pkg{mgcViz} package \cite{mgcvizpackage}. In the following we will only use the \pkg{mgcViz} tools that are specific to quantile GAMs, and we refer to \cite{fasiolo2018scalable} and to the relevant package documentation for more details. To exploit the visualizations offered by \pkg{mgcViz}, it is necessary to transform the output of \code{mqgam} as follows 
\begin{Schunk}
\begin{Sinput}
R> library("mgcViz")
R> fitCycleM <- getViz(fitCycleM)
\end{Sinput}
\end{Schunk}
which produces an object of class \code{c("mqgamViz", "mgamViz")}. This is simply a list of objects of class \code{"gamViz"}, which inherits from the \code{"qgam"} class. Given that the elements of \code{fitCycleM} are full objects, without any missing slot, they can be manipulated directly rather than via \code{qdo}. For example, we can use \code{plot(fitCycleM[[1]])} to plot the fitted effect for quantile $\tau = 0.1$. Of course, transforming \code{fitCycleM} using \code{getViz} has some memory cost, because \code{fitCycleM} now stores a copy of the model matrix for each quantile. 

Having transformed \code{fitCycleM}, the effect of time on all quantiles can now be visualized jointly by
\begin{Schunk}
\begin{Sinput}
R> plot(fitCycleM)
\end{Sinput}
\end{Schunk}
\begin{figure}[t]
\centering
\begin{Schunk}

\includegraphics[width=\maxwidth]{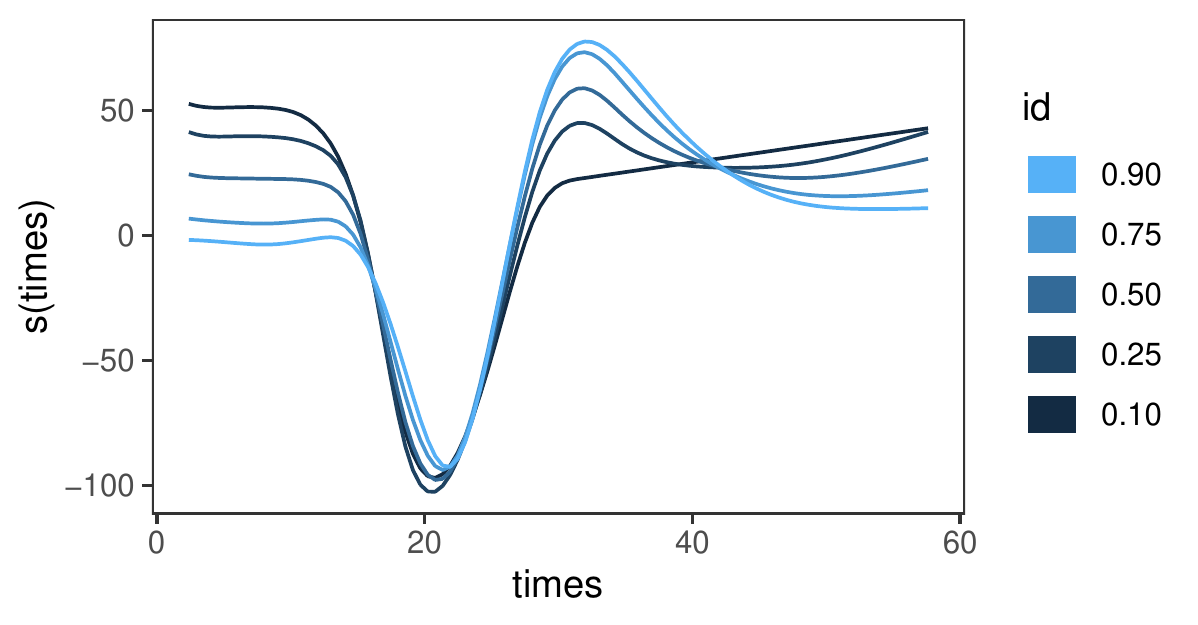} \end{Schunk}
\caption{\label{fig:mcyclemgcViz} Plot showing the fitted additive effects of time on all five quantiles, obtained using the \pkg{mgcViz} package.}
\end{figure}
which produces the plot shown in Figure \ref{fig:mcyclemgcViz}, by dispatching to \code{mgcViz::plot.mgamViz}. The latter produces an object of class \code{c("plotGam", "gg")}, which is a wrapper around one or several objects of class \code{ggplot}, defined in the \pkg{ggplot2} package \citep{ggplot2package}.  Note that Figure \ref{fig:mcyclemgcViz} shows the fitted effects for each $\tau$ which, in this case, are equal to the fitted quantiles up to some vertical shifts (the intercepts). The effects cross each other because they are all centered (see \cite{wood2006generalized} for details on identifiability constraints in GAMs), hence this does not imply that the corresponding quantile estimates (which are not centered) cross within the range of the data, $\text{time} \in (0, 60)$. In fact, if we plot the estimated quantiles using 
\begin{Schunk}
\begin{Sinput}
R> xseq <- with(mcycle, seq(min(times), max(times), length.out = 100))
R> preds <- sapply(fitCycleM, predict, newdata = data.frame(times = xseq))
R> plot(mcycle, ylim = range(preds))
R> for(ii in 1:5) lines(xseq, preds[ , ii], col = 2)
\end{Sinput}
\end{Schunk}
\begin{figure}[t]
\centering
\begin{Schunk}

\includegraphics[width=\maxwidth]{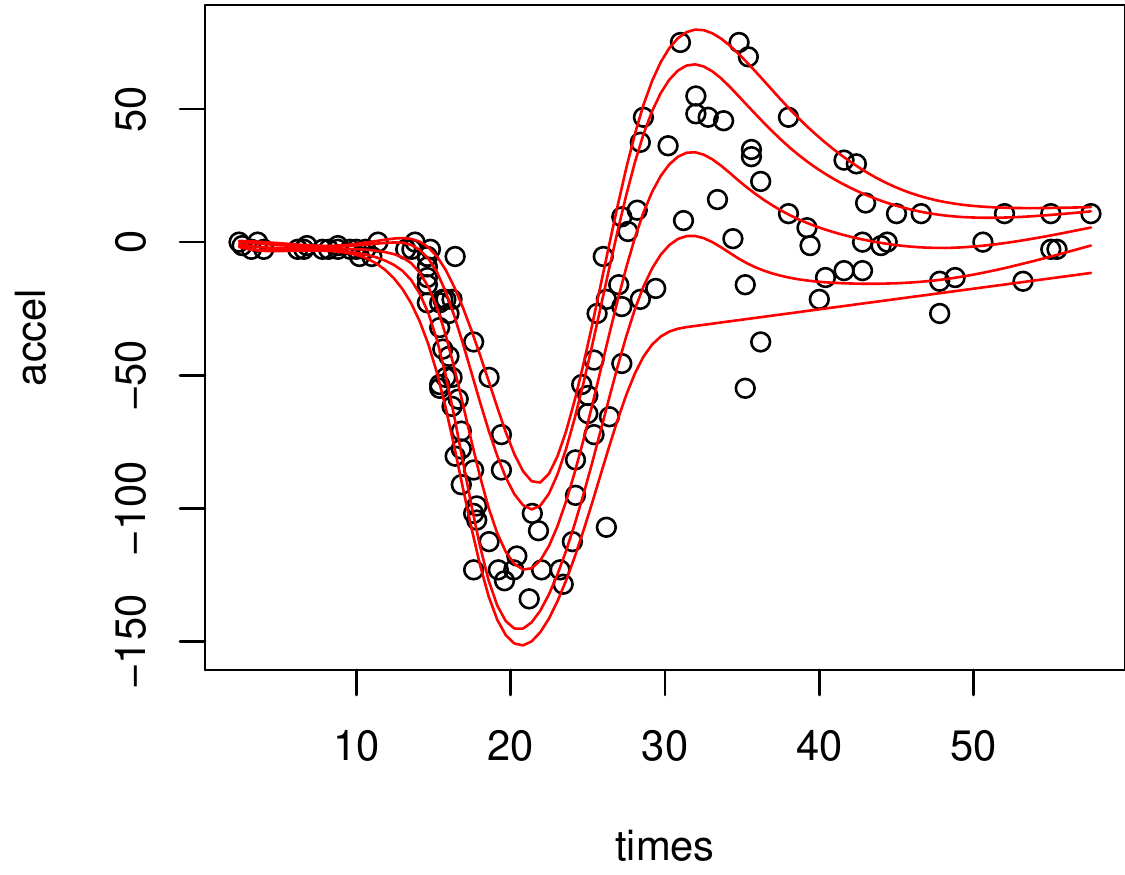} \end{Schunk}
\caption{\label{fig:mcycleWithInterc} Plot showing the five conditional quantiles fitted using \code{qgam}.}
\end{figure}
we do not see any clear crossing in Figure \ref{fig:mcycleWithInterc}. Indeed, the minimal distance between consecutive quantiles is
\begin{Schunk}
\begin{Sinput}
R> min(apply(preds, 1, diff))
\end{Sinput}
\begin{Soutput}
[1] 0.1524413
\end{Soutput}
\end{Schunk}
hence there is no crossing within the range of the data. However, quantiles fitted with \pkg{qgam} often cross somewhere (outside the data range in this example), because they are fitted independently and without any non-crossing constraint. This issue is not specific to \pkg{qgam}, but is a well known drawback of distribution-free quantile regression \citep{koenker2005quantile}. 

\begin{figure}
\centering
\begin{Schunk}

\includegraphics[width=\maxwidth]{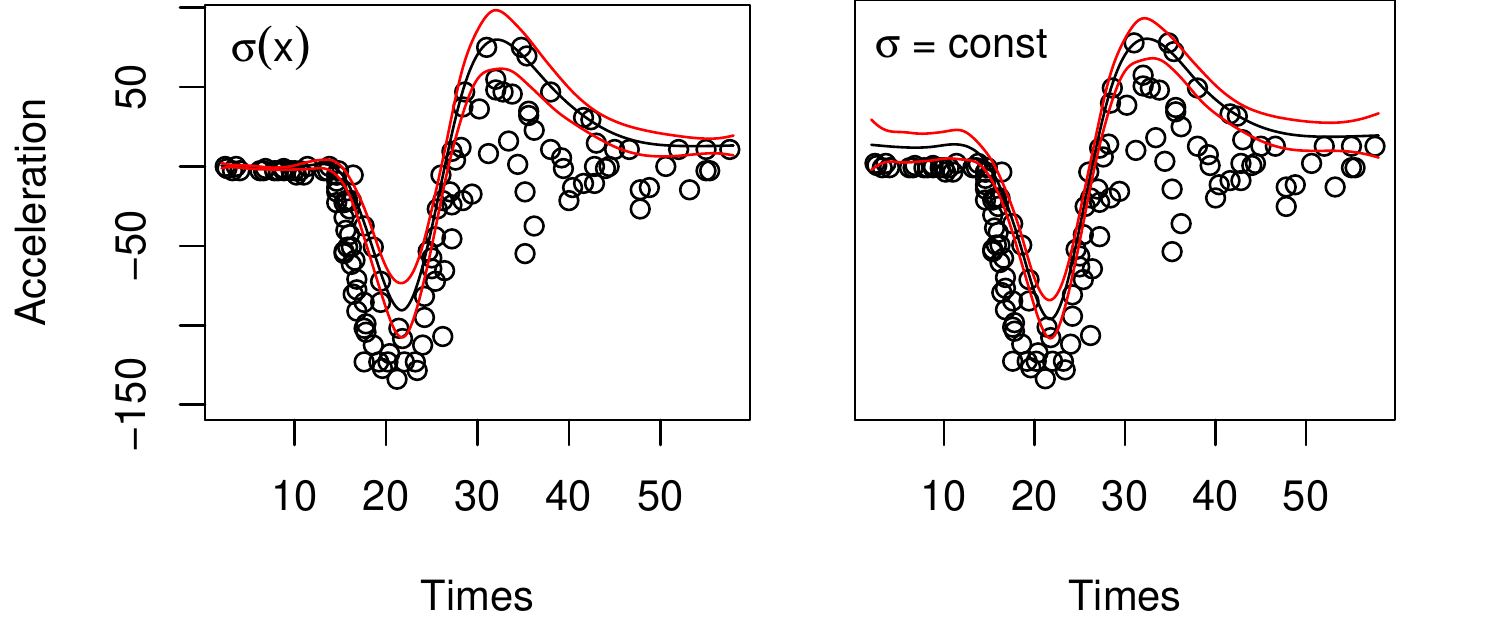} \end{Schunk}
\caption{\label{fig:mcycleNoVar} Fitted quantile $\tau = 0.9$, with 95\% credible intervals, when the learning rate and the loss smoothness are allowed to vary with the independent variable (left) or not (right).}
\end{figure}

In the \code{mcycle} data set, the variance of the response (\code{accel}) varies strongly with the independent variable (\code{times}), hence it is natural to let the learning rate and the ELF loss smoothness vary along the latter. This is achieved by providing a model formula which is a list with two elements, where the second, \code{~ s(times)}, is used by \code{qgam} to model the conditional variance $\kappa(\bm x)$ in the location-scale GAM model. However, it is possible to avoid modelling the variance and to assume that it is constant along the covariates. For example, in the following code
\begin{Schunk}
\begin{Sinput}
R> fitCycleConst <- qgam(accel ~ s(times, k = 20, bs = "ad"), 
+    data = mcycle, qu = 0.9)
\end{Sinput}
\end{Schunk}
we are providing only the model for the quantile $\tau = 0.9$, rather than a list of two formulas, so \code{qgam} uses a constant learning rate $1/\sigma(\bm x) = 1/\sigma_0$ and loss bandwidth $\tilde{h}^*(\bm x) = \tilde{h}^*$. For the \code{mcycle} data set, ignoring the heteroscedasticity leads to the poor results shown in Figure \ref{fig:mcycleNoVar}. 
One problem with the quantile fit shown on the right is that it lies far above all the responses, for $\text{time} < 10$ms. But we are fitting quantile 0.9, hence we should expect around 10$\%$ of the responses to lie above the fit. The issue is that the bias of the ELF loss used by \code{qgam} is inversely proportional to the conditional variance of the response (see \cite{fasiolo2017fast} for details), hence not modelling the response variance leads to high bias for $\text{time} < 10$ms. Ignoring heteroscedasticity can also lead to credible intervals that have non-constant coverage. In fact, the width of the intervals should be proportional to the variance of the response, but on the right plot in Figure \ref{fig:mcycleNoVar} they have nearly constant width, thus providing an incorrect representation of the uncertainty of the fit. Of course, in most data set, heteroscedasticity is not as dramatic as in \code{mcycle} and using a location-only model often leads to satisfactory results. However, it is important to be aware of the issues highlighted by this example.   

\subsection{Functional QGAM modelling of electricity demand data} \label{sec:electDemandExam}

In this example we consider smart meter data from $n_c = 247$ anonymized residential customers from Sydney, Australia, covering the period between the 3rd of July 2010 and the 30th of June 2011. The data has been downloaded from \url{https://www.ausgrid.com.au}, and it originally contained electricity demand from 300 customers, at 30min resolution. We discarded 53 customers because their demand was too irregular (e.g., they were absent from home for over 30 consecutive days), and we integrated the demand data with temperature data from the National Climatic Data Center (\url{www.ncdc.noaa.gov}), covering the same period. The aim of this example is to illustrate how to use quantile GAMs in \pkg{qgam}, in the context of an electricity demand forecasting application. We start by loading the following data set:
\begin{Schunk}
\begin{Sinput}
R> data("AUDem", package = "qgam")
R> meanDem <- AUDem$meanDem
R> head(meanDem, 3)
\end{Sinput}
\begin{Soutput}
  doy  tod       dem dow     temp                date     dem48
1 184 18.0 0.8248777 Sat 3.357407 2010-07-03 17:30:00 0.8978636
2 184 18.5 0.8686110 Sat 2.517073 2010-07-03 18:00:00 0.9417633
3 184 19.0 0.8519471 Sat 1.898399 2010-07-03 18:30:00 0.9148921
\end{Soutput}
\end{Schunk}
which contains the variables:
\begin{itemize}
\item \code{doy} the day of the year, from 1 to 365;
\item \code{tod} the time of day, ranging from 18 to 22, where 18 indicates
                 the period from 17:00 to 17:30, 18.5 the period from 17:30 to 18:00 and so on;
\item \code{dem} the demand (in KWh) during a 30min period, averaged over the $n_c$ households;  
\item \code{dow} factor variable indicating the day of the week;
\item \code{temp} the external temperature at Sydney airport, in degrees Celsius;
\item \code{date} local date and time;
\item \code{dem48} the lagged mean demand, that is the average demand (\code{dem}) during the same 30min period of the previous day;
\end{itemize}

\begin{figure}[t]
\centering
\begin{Schunk}

\includegraphics[width=\maxwidth]{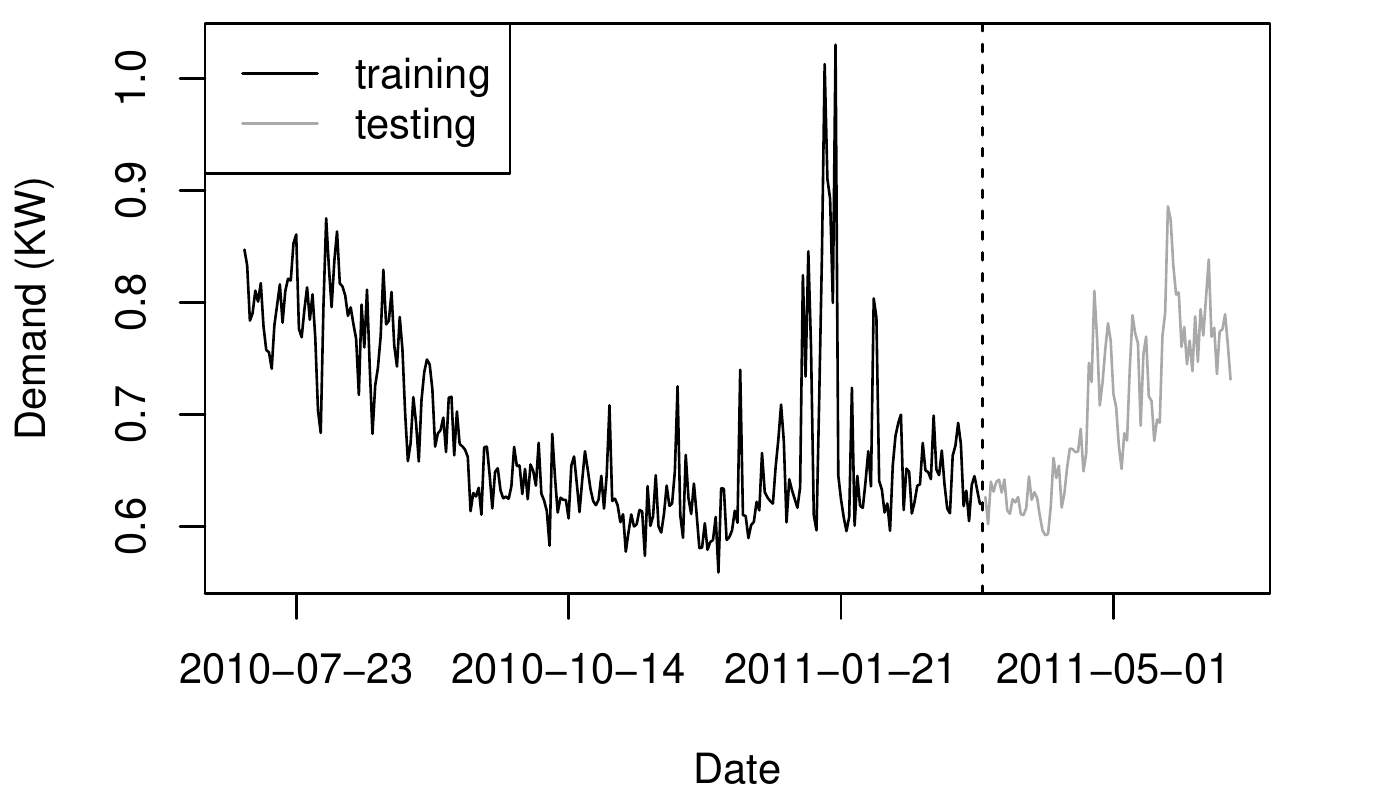} \end{Schunk}
\caption{\label{fig:electDem} Electricity demand between 19:30 and 20:00, averaged across the $n_c$ customers.}
\end{figure}

Assume that we aim at producing a probabilistic forecast of the average demand one day ahead, using additive quantile regression. We start by dividing the data into a learning and a testing set:
\begin{Schunk}
\begin{Sinput}
R> cutDate <- as.Date("2011-04-01 00:00:00")
R> meanDemLearn <- subset(meanDem, as.Date(date) < cutDate)    # Learning
R> meanDemTest <- subset(meanDem, as.Date(date) >= cutDate)    # Testing
\end{Sinput}
\end{Schunk}
which leaves the last three months for testing. Figure \ref{fig:electDem} shows the average demand between 19:30 and 20:00, over the whole period. We focus only on the period between 17:30 and 21:30, because the demand dynamics change considerably before and after this time slot, hence modelling demand across the whole day would require a much more sophisticated and computationally expensive model than any of those considered below.

We start by modelling the $\tau$-th quantile of the average demand using the quantile model
\begin{equation} \label{eq:electMod1}
\mu_\tau({\bm x}_t) = \sum_{j=1}^7\beta_j\mathbb{I}(\text{dow}_t = j) + \beta_8\text{dem}_{t-48} + f_1(\text{tod}_t) + f_2(\text{temp}_t) + f_3(\text{doy}_t),
\end{equation}
where $\mathbb{I}(\text{dow}_t = j) = 1$ when $\text{dow}_t$ is the $j$-th day of the week and zero otherwise, $f_1$ and $f_2$ are smooth effects built using thin plate splines, while $f_3$ is a cyclical effect constructed using cubic regression splines. We use a cyclic effect for \code{doy} to make so that the seasonal effect has the same value on the 31st of December and on the 1st of January. Given that demand variability seems to spike during the austral summer (see Figure \ref{fig:electDem}), we model the variance $\kappa(\bm x)$ using a cyclic effect for \code{doy} in the location-scale Gaussian GAM. 

We fit the model just described to quantiles $\tau = 0.1, 0.3, 0.5, 0.7, 0.9$ using the following code
\begin{Schunk}
\begin{Sinput}
R> library("qgam")
R> qusObj <- seq(0.1, 0.9, length.out = 5) 
R> fitQ <- mqgam(list(dem ~ dow + dem48 + 
+    s(tod, k = 6) + s(temp) + s(doy, bs = "cc"), ~ s(doy, bs = "cc")), 
+    qu = qusObj, data = meanDemLearn) 
\end{Sinput}
\end{Schunk}
We then plot the estimated smooth effects and the linear coefficients of lagged demand using 
\begin{Schunk}
\begin{Sinput}
R> library("mgcViz")
R> fitQ <- getViz(fitQ)
R> print(plot(fitQ, allTerms = TRUE, select = c(1:3, 5)), pages = 1)
\end{Sinput}
\end{Schunk}
\begin{figure}[t]
\centering
\begin{Schunk}

\includegraphics[width=\maxwidth]{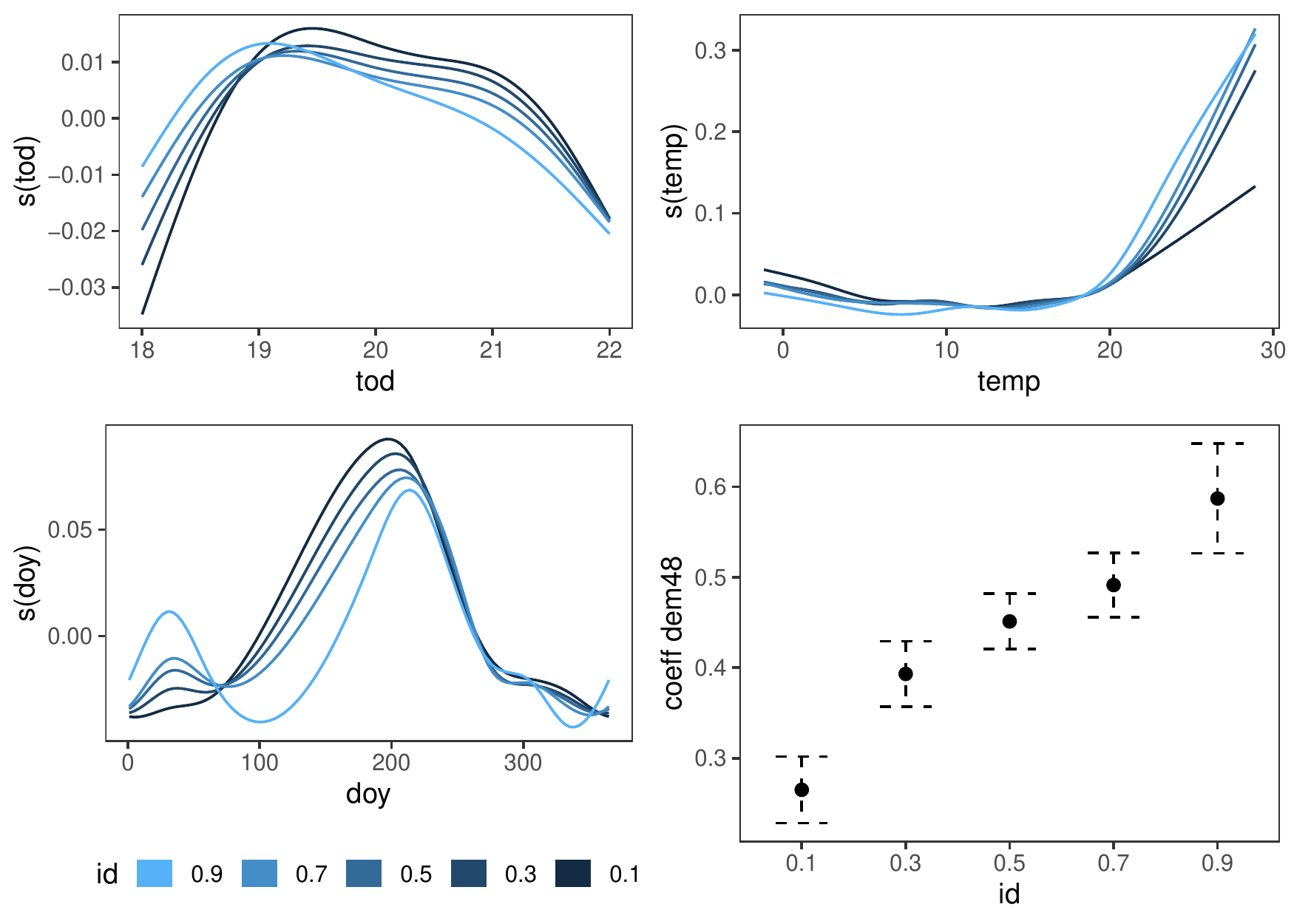} \end{Schunk}
\caption{\label{fig:elecQGAM1} Estimated effects of time of day (\code{tod}), temperature (\code{temp}), day of year (\code{doy}) and lagged demand (\code{dem48}) from model (\ref{eq:electMod1}), for the quantiles labelled using the \code{id} variable.}
\end{figure}
The output, up to some aesthetic manipulation, is shown in Figure \ref{fig:elecQGAM1}. Note that we used the plotting tools provided by \code{mgcViz}, which required us to convert the output of \code{mqgam} using \code{getViz}. Then we used \code{plot} to plot all model terms (\code{allTerms = TRUE} is used to plot the parametric terms in addition to the smooth effects), we selected a subset of the effects using the \code{select} argument and we used \code{print} to arrange all the plots on a single page.

The shapes of the estimated effects seem quite reasonable: the \code{tod} effect increases as people get home and decreases later in the evening, the temperature plot shows a strong cooling effect and a negligible heating effect, and the effect of \code{doy} shows a main mode during the austral winter and a lower, narrower, mode during the summer. The discrepancy between the effects estimated on different quantiles can be interpreted in terms of the corresponding effect on the shape of the demand distribution. For example, at \code{tod} = 18 the effects $f_1$ are more spread out than at \code{tod} = 19 implying that, all else being equal, the variance of the demand is higher at \code{tod} = 18. Similarly, at \code{doy} $\approx 25$ the effect $f_3$ is more positive for quantile $\tau = 0.9$ than for $\tau = 0.1$, while the positions are reversed at \code{doy} $\approx 120$. Hence, the demand distribution is more skewed to the right in the first case. The linear coefficient of the lagged demand increases with $\tau$ hence, all else being equal, the demand variance one day ahead increases with the lagged demand.     

So far we considered only the mean demand over the portfolio of customers, however the original data set contains demand for each of the $n_c$ households. Average demand is typically much easier to forecast than individual demand, because the highly irregular consumption of individual customers is averaged out. However, the by-customer data should carry more information, hence it is interesting to verify whether it can be used to improve the quantile forecasts for the average demand. We capture part of the information contained in the by-customer demand data by considering a model which, for storage reasons, uses a functional summary of the full demand data, as explained in the following. 

We consider the functional quantile model
\begin{equation} \label{eq:electMod2}
\mu_\tau({\bm x}_t) = \sum_{j=1}^7\beta_j\mathbb{I}(\text{dow}_t = j) +  f_1(\text{tod}_t) + f_2(\text{temp}_t) + f_3(\text{doy}_t) + \int_0^{1} f_4(p)G^{-1}_{t-48}(p)dp,
\end{equation}
where $\mu_\tau({\bm x}_t)$ is, as in (\ref{eq:electMod1}), the $\tau$-th quantile of the average demand, while we indicate with $G_t$ the c.d.f. of the individual demand across customers at time $t$ and with $G^{-1}_{t-48}(p)$ the corresponding $p$-th quantile, at the same instant of the previous day. Function $f_4$ is a smooth effect along $p \in [0, 1]$, constructed using a thin plate spline basis. Hence, the last term in (\ref{eq:electMod2}) is a functional effect which takes as input the whole quantile function of the across-customers demand distribution at time $t-48$, and integrates it over $p$ using the weight function $f_4(p)$. The aim is to verify whether the shape of the distribution of the individual demand at time $t-48$ can be used to improve the forecast of the average demand at time $t$. The regression coefficients of $f_4(p)$ are unknown, but they can be estimated using the methods presented in Section \ref{sec:modelFitting}. In particular, we have that  
$$
\int_0^{1} f_4(p)G^{-1}_{t-48}(p)dp = \int_0^{1} \sum_{k=1}^{K_4}b_{k}^4(p)\beta_k^4G^{-1}_{t-48}(p)dp =  \sum_{k=1}^{K_4}\left\{\int_0^{1} b_{k}^4(p)G^{-1}_{t-48}(p)dp\right\}\beta_k^4 = \sum_{k=1}^{K_4}\tilde{b}_{k}^4\beta_k^4,
$$
where the transformed basis $\tilde{b}_{1}^4, \dots, \tilde{b}_{K_4}^4$ implicitly depends on time $t$. Hence, once the transformed basis is obtained, the functional effect is linear in the regression coefficients and estimating them presents no extra difficulty, relative to standard smooth effects. The across-customers demand c.d.f. $G_{t-48}$ is unknown, but we have a sample of size $n_c$ from it, consisting of the by-customer demand data. Hence, we can approximate the transformed basis using
\begin{equation} \label{eq:approxBasis}
\tilde{b}_{k}^4 = \int_0^{1} b_{k}^4(p)G^{-1}_{t-48}(p)dp \approx \frac{1}{n_q}\sum_{l=1}^{n_q} \ b_{k}^4\left(p_l\right) \hat{G}_{t-48}^{-1}(p_l),
\end{equation}
where $\hat{G}_{t-48}^{-1}(p_1), \dots, \hat{G}_{t-48}^{-1}(p_{n_q})$ is a set of empirical quantiles of the $n_c$ by-customer demand observations at time $t-48$, corresponding to probability level $0<p_1<p_2<\cdots<p_{n_q}<1$. 

\begin{figure}
\centering
\begin{Schunk}

\includegraphics[width=\maxwidth]{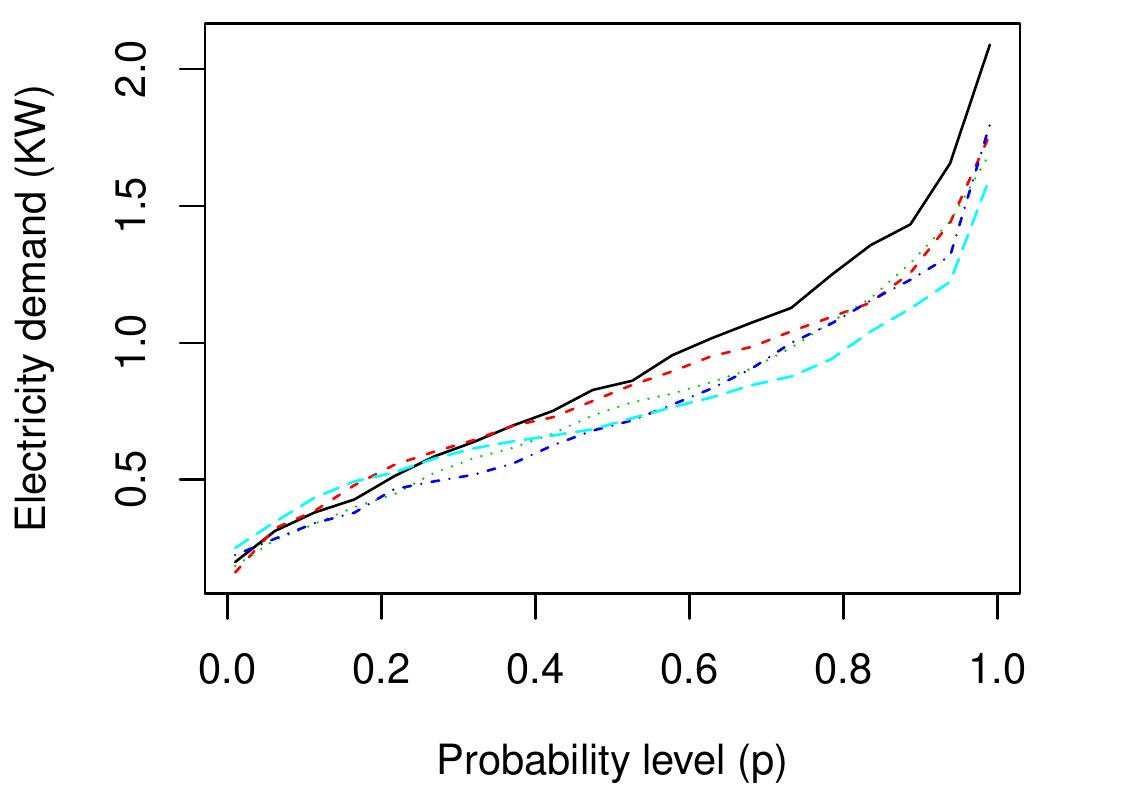} \end{Schunk}
\caption{\label{fig:QuantFunc} Estimates of the individual demand quantile function $G^{-1}_{t-48}$, for five values of $t$.}
\end{figure}

To reduce the amount of data stored in the \pkg{qgam} package, we use $n_q = 20 < n_c$ quantiles equally spaced between $p_1 = 0.01$ and $p_{n_q} = 0.99$. The quantiles are stored in 
\begin{Schunk}
\begin{Sinput}
R> qDem48 <- AUDem$qDem48
R> qDem48[1:3, c(1, 2, 5, 15, 19, 20)]
\end{Sinput}
\begin{Soutput}
            1% 6.157895% 21.63158% 73.21053% 93.84211%      99%
[1,] 0.1995373 0.3128564 0.5125927  1.127776  1.655850 2.089360
[2,] 0.1595543 0.3368127 0.5569682  1.206294  1.587732 2.084527
[3,] 0.2091480 0.3571971 0.5652339  1.141480  1.632144 2.044097
\end{Soutput}
\end{Schunk}
where each row contains $\hat{G}_{t-48}^{-1}(p_1), \dots, \hat{G}_{t-48}^{-1}(p_{n_q})$, for some $t$. Figure \ref{fig:QuantFunc} shows five selected rows of \code{qDem48}, each line being an estimate of the quantile function $G^{-1}_{t-48}$, for a different $t$. 
To fit model (\ref{eq:electMod2}) with \pkg{qgam}, we need also a matrix where each row of \code{probLev} contains the probability levels $p_1, p_2, \dots, p_{n_q}$. This is computed as follows
\begin{Schunk}
\begin{Sinput}
R> probLev <- matrix(seq(0.01, 0.99, length.out = 20), 
+    nrow = nrow(meanDem), ncol = 20, byrow = TRUE)
R> probLev[1:3, c(1, 2, 5, 15, 19, 20)]
\end{Sinput}
\begin{Soutput}
     [,1]       [,2]      [,3]      [,4]      [,5] [,6]
[1,] 0.01 0.06157895 0.2163158 0.7321053 0.9384211 0.99
[2,] 0.01 0.06157895 0.2163158 0.7321053 0.9384211 0.99
[3,] 0.01 0.06157895 0.2163158 0.7321053 0.9384211 0.99
\end{Soutput}
\end{Schunk}
We add the matrices just defined to the learning and testing sets by
\begin{Schunk}
\begin{Sinput}
R> ntrain <- nrow(meanDemLearn)
R> ntest <- nrow(meanDemTest)
R> meanDemLearn$qDem48 <- qDem48[1:ntrain, ]
R> meanDemTest$qDem48  <- qDem48[-(1:ntrain), ]
R> meanDemLearn$probLev <- probLev[1:ntrain, ]
R> meanDemTest$probLev <- probLev[-(1:ntrain), ]
\end{Sinput}
\end{Schunk}
We are now ready to fit model (\ref{eq:electMod2}) using
\begin{Schunk}
\begin{Sinput}
R> fitFunQ <- mqgamV(list(dem ~ dow + s(temp) + s(doy, bs = "cc") + 
+    s(tod, k = 6) + s(probLev, by = qDem48), ~ s(doy, bs = "cc")),
+    qu = qusObj, data = meanDemLearn)
\end{Sinput}
\end{Schunk}
where we are using the \code{mgcViz::mqgamV} function, which is simply a shortcut for fitting the model using \code{mqgam} and then transforming it using \code{getViz}. Here the functional effect is constructed using \code{s(probLev, by = orDem48)}, which is the constructor typically used for by-variable or varying-coefficient effects in \pkg{mgcv}. The corresponding transformed spline basis is constructed by \code{gam}, which performs the summation in (\ref{eq:approxBasis}) using the rows of the \code{probLev} and \code{orDem48} matrices. 

\begin{figure}
\centering
\begin{Schunk}

\includegraphics[width=\maxwidth]{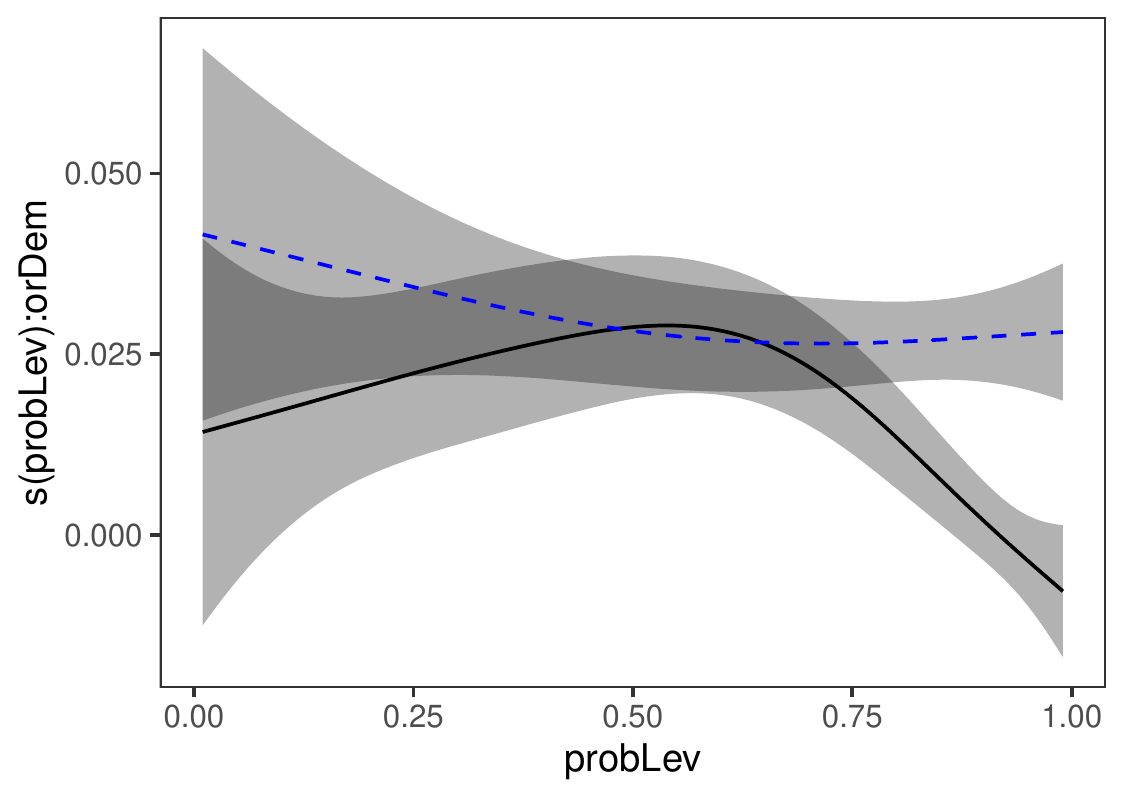} \end{Schunk}
\caption{\label{fig:plotFunEffCompare} Estimated effect $f_4(p)$ from model (\ref{eq:electMod2}) with 95\% credible intervals, for quantile $\tau = 0.1$ (black solid) and 0.9 (blue dashed).}
\end{figure}

We can have a look at the fitted functional smooth effects $f_4(p)$ using
\begin{Schunk}
\begin{Sinput}
R> plot(fitFunQ, select = 4) + labs(color = expression(tau))
\end{Sinput}
\end{Schunk}
\begin{figure}
\centering
\begin{Schunk}

\includegraphics[width=\maxwidth]{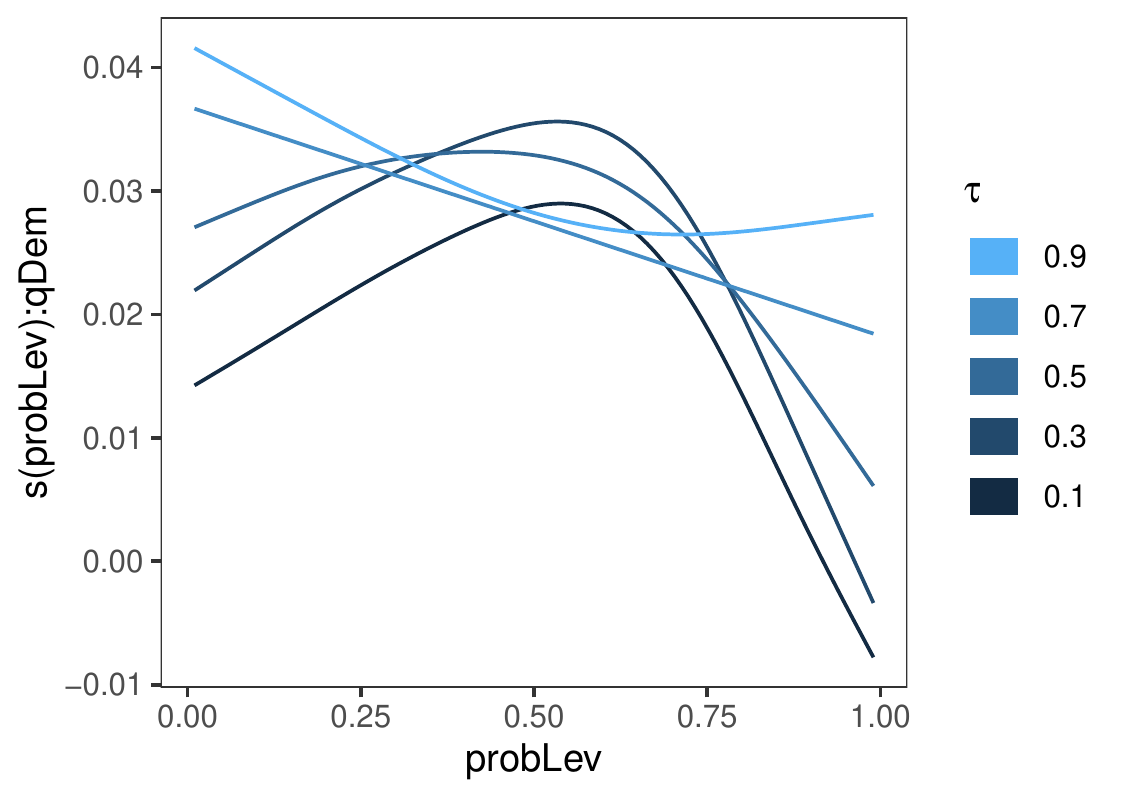} \end{Schunk}
\caption{\label{fig:plotFunEff} Estimated effect $f_4(p)$ from model (\ref{eq:electMod2}), for each quantile.}
\end{figure}
which produces the plot shown in Figure \ref{fig:plotFunEff}. It is interesting to note that the effects corresponding to low and high probabilities diverge, suggesting that the extremes of the quantile function $G^{-1}_{t-48}(p)$ have a stronger linear effect on the high quantiles of the average demand (e.g., $\mu_\tau(\bm x)$ with $\tau = 0.9$) than on the low ones. However, the \code{plot.mgamViz} method does not include credible intervals when plotting the same effect for multiple quantiles, to avoid cluttering the plot. In Figure \ref{fig:plotFunEffCompare} we compare the effects for $\tau = 0.1$ and 0.9, and we include 95\% credible intervals. This shows that the discrepancy between the effects is statistically stronger for $p \approx 1$, than for $p \approx 0$. 

To check whether model (\ref{eq:electMod2}) leads to a better fit than model (\ref{eq:electMod1}), we can compare the values of the Bayesian information criterion (BIC) corresponding to each quantile by
\begin{Schunk}
\begin{Sinput}
R> round(rbind(QGAM = sapply(fitQ, BIC), FuncQGAM = sapply(fitFunQ, BIC)))
\end{Sinput}
\begin{Soutput}
           0.1   0.3   0.5   0.7   0.9
QGAM     -8255 -9513 -9700 -9137 -6651
FuncQGAM -8458 -9617 -9698 -9152 -6680
\end{Soutput}
\end{Schunk}
BIC favours slightly model (\ref{eq:electMod2}), but the simpler model is preferred for one of the quantiles. Instead the Akaike information criterion (AIC), which penalizes model complexity less strongly than BIC, favours model (\ref{eq:electMod2}) for all quantiles
\begin{Schunk}
\begin{Sinput}
R> round(rbind(QGAM = sapply(fitQ, AIC), FuncQGAM = sapply(fitFunQ, AIC)))
\end{Sinput}
\begin{Soutput}
           0.1   0.3   0.5   0.7   0.9
QGAM     -8416 -9679 -9867 -9303 -6813
FuncQGAM -8637 -9800 -9881 -9325 -6858
\end{Soutput}
\end{Schunk}
We can also compare the predictive performance of the two models on the test set, which comprises the last three months of data set (1st of April to 30th of June, see Figure \ref{fig:electDem}).
For example, we can calculate the pinball loss for each model and quantile on such data as follows
\begin{Schunk}
\begin{Sinput}
R> predQ <- sapply(fitQ, predict, newdata = meanDemTest)
R> predFunQ <- sapply(fitFunQ, predict, newdata = meanDemTest)
R> round(rbind(QGAM = pinLoss(meanDemTest$dem, predQ, qusObj), 
+    FuncQGAM = pinLoss(meanDemTest$dem, predFunQ, qusObj)), 1)
\end{Sinput}
\begin{Soutput}
         0.1 0.3  0.5  0.7  0.9
QGAM     4.7 8.8 10.4 10.6 10.7
FuncQGAM 4.6 8.6 10.3 10.2 10.2
\end{Soutput}
\end{Schunk}
where \code{qgam::pinLoss} evaluates the pinball loss. The functional model does better on the test set, but it is important to point out that the size of the test set is rather limited, especially for the purpose of evaluating the performance of the extreme quantiles estimates.

Of course, model (\ref{eq:electMod2}) could be improved further. For example, a call to
\begin{Schunk}
\begin{Sinput}
R> check(fitFunQ[[3]])
\end{Sinput}
\begin{Soutput}
Theor. proportion of neg. resid.: 0.5   Actual proportion: 0.5155229
Integrated absolute bias |F(mu) - F(mu0)| = 0.02719397
Method: REML   Optimizer: outer newton
full convergence after 7 iterations.
Gradient range [-3.421902e-08,4.411971e-07]
(score -4857.326 & scale 1).
Hessian positive definite, eigenvalue range [0.5027064,3.678087].
Model rank =  39 / 39 

Basis dimension (k) check: if edf is close too k' (maximum possible edf) 
it might be worth increasing k. 

                  k'  edf
s(temp)            9 8.21
s(doy)             8 7.65
s(tod)             5 4.69
s(probLev):qDem48 10 3.15
\end{Soutput}
\end{Schunk}
shows that, for $\tau = 0.5$, the effective degrees of freedom (\code{edf}) are quite close to their maximum possible value (\code{k'}) for the temperature effect. This means that the fit is using all the degrees of freedom available for that effect, hence we might want to consider increasing the number of basis functions by using \code{s(temp, k = 20)} in the model formula. However, one might be reluctant to increase the number of basis functions used to fit more extreme quantiles (e.g., $\tau = 0.9$), where the data is sparser. This highlights a disadvantage of using \code{mqgam}, rather than repeated calls to \code{qgam}, when fitting several QGAMs: the same model formula is used for all the quantiles. The model could also be improved by, for example, including the effect of lagged temperatures, often useful for capturing thermal inertia in buildings.   

\section{Conclusions} \label{sec:conclusions}

The \pkg{qgam} \proglang{R} package provides tools for building, fitting and checking quantile additive models. The main advantage of such models, relative to standard probabilistic GAMs, is that they do not make any parametric assumption on the distribution of the response variable. This is achieved by modelling and estimating the conditional quantiles directly, using the pinball loss. Model fitting is based on the fast calibrated Bayesian framework of \cite{fasiolo2017fast}, which provides calibrated credible intervals, via IKL selection of the learning rate, and computational efficiency, by smoothing the pinball loss and by exploiting the marginal likelihood methods of \cite{wood2016smoothing}. Model building is handled by \pkg{mgcv}, hence the quantile GAMs built using \pkg{qgam} can include any of the effect types made available by \pkg{mgcv}, among others: cyclic, adaptive and multivariate tensor-product smooths, as well as functional and random effects. 

At the time of writing, \pkg{qgam} version \code{1.3.2} can handle data sets and models of moderate size. In particular, the current fitting methods require explicit formation of the $n \times d$ model matrix $\bf X$, which leads to high memory usage for large $n$. \cite{wood2015generalized} proposed a GAM fitting framework where explicit formation of $\bf X$ is avoided by adopting block-oriented iterations for smoothing parameters selection and regression coefficients estimation. The \code{bam} function in \pkg{mgcv} implements such methods, as well as the marginal discretization approach of \cite{wood2017generalized}, which allows it to fit models with up to $10^4$ coefficients and $10^8$ data points. Future work will aim at improving the scalability of the Bayesian QGAM fitting framework currently implemented in \pkg{qgam}, by exploiting the block-oriented discretized fitting methods provided by \pkg{mgcv} for the inner and intermediate iterations, and developing a similarly scalable version of the IKL minimization routine currently used for learning rate selection.

\section*{Acknowledgments}

The development of the \pkg{qgam} package was funded by EPSRC grants EP/K005251/1, EP/N509\\619/1 and by {\'E}lectricit{\'e} de France. M. Zaffran gratefully acknowledges support from the Erasmus+ programme and the Universit{\'e} Paris-Saclay.


\bibliography{refs}

\newpage

\begin{appendix}

\section{Choosing the loss bandwidth manually} \label{sec:lossBand}

The \code{qgam} function determines the ELF loss bandwidth $h(\bm x)$ automatically, but this parameter can be chosen manuallly by specifying the \code{err} argument, which is set to \code{NULL} by default. Here we provide some guidelines for users who wish to choose this parameter manually. In particular, the examples in Section \ref{sec:exErr} illustrate how parameter \code{err} affects the statistical and computational performance of the fitting methods provided by \pkg{qgam}.

\subsection{Illustrating the effect of the loss bandwidth and some diagnostics} \label{sec:exErr}

To clarify the interpretation of parameter \code{err}, let us consider a univariate context where we want to estimate the $\tau$-th quantile of response $y$ and there are no covariates. Define $\epsilon = |F(\mu^*) - F(\mu_0)|$, where $F$ is the c.d.f. of $y$, $\mu_0$ is the true quantile at probability level $\tau$ and $\mu^*$ is the minimizer of $\mathbb{E}\{\tilde{\rho}_\tau(y-\mu)\}$. Given that $\mu_0$ is the minimizer of the expected pinball loss (see (\ref{eq:quantDef})), $\epsilon$ is the asymptotic absolute bias induced by the use of the smooth ELF loss, with bandwidth $h = \lambda \sigma$. In Appendix \ref{app:approxQualProof} we prove that 
\begin{equation} \label{eq:biasBound}
\epsilon = |F(\mu^*) - F(\mu_0)| \leq 2 \log2 \, h \, \underset{y}{\text{sup}} f(y) = \frac{ 2 \log2}{\sqrt{2 \pi \kappa}} \, h,   
\end{equation}
where $f$ is the p.d.f. of $y$ and the last equality holds if $y$ is Gaussian with variance $\kappa$. Via argument \code{err}, the \code{qgam} function allows users to set the maximum tolerable bias $\epsilon$, which is then used to determine the loss bandwidth by 
\begin{equation} \label{eq:manualErr}
h = \epsilon \frac{\sqrt{2\pi\kappa}}{2 \log 2}.
\end{equation}
Here $\kappa$ is estimated via the location-scale GAM model, and can vary with $\bm x$ as usual. Parameter $h = h(\bm x)$, obtained via (\ref{eq:manualErr}), is then used to determine $\lambda$ and $\sigma(\bm x)$ as in Section \ref{sec:modelFitting}. Of course $\epsilon$ (\code{err} in the software) is an approximate bound, because the response is not normally distributed in practice. 

In \pkg{qgam} we let users choose $\epsilon$, rather than $\lambda$, because it does not depend on the scale of the response $y$. Indeed, considering the interpretation of $\epsilon$, reasonable values of this parameter fall in $(0, 1)$ (but \code{qgam} checks only that \code{err} is positive). To illustrate how \code{err} influences the accuracy of the quantile estimates and the computational performance of \code{qgam}, we consider a simple data set simulated as follows 
\begin{Schunk}
\begin{Sinput}
R> set.seed(5523)
R> x <- seq(-3, 3, length.out = 1e3)
R> X <- data.frame("x" = x, "y" = x + x^2 + rgamma(1e3, 4, 1))
\end{Sinput}
\end{Schunk}
Hence $y_i = x_i + x_i^2 + z_i$, where $z_i \sim \text{Gamma}(4, 1)$. Assuming that we want to estimate quantiles 0.05, 0.5 and 0.95, the following code estimates them using different values of \code{err}
\begin{Schunk}
\begin{Sinput}
R> fitGrid <- lapply(c(0.01, 0.05, 0.1, 0.3, 0.5),
+    function(.errVal){
+    mqgam(y ~ s(x), data = X, qu = c(0.05, 0.5, 0.95), err = .errVal) 
+    })
\end{Sinput}
\end{Schunk}
\begin{figure}[t]
\centering
\begin{Schunk}

\includegraphics[width=\maxwidth]{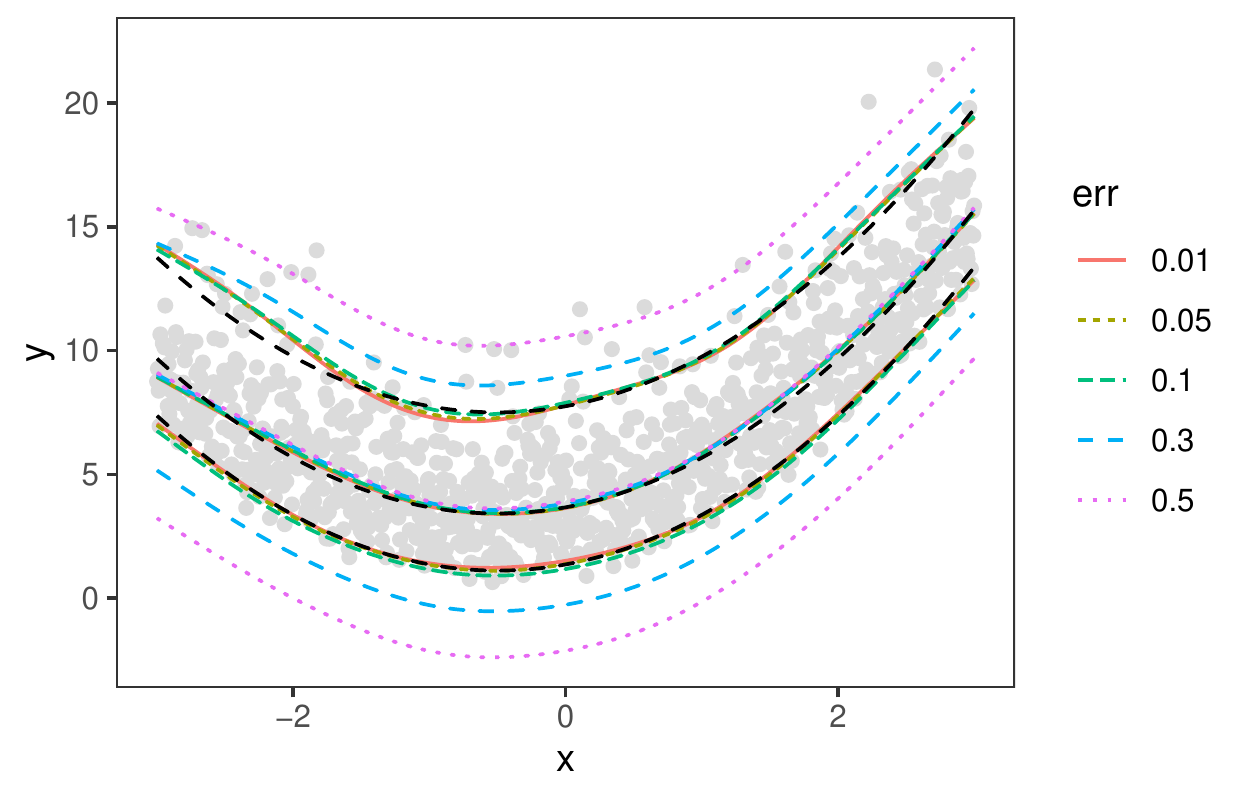} \end{Schunk}
\caption{\label{fig:plotErrLev} Quantile fits for $\tau = 0.05, 0.5$ and 0.95 for different values of \code{err}. The dashed lines are the true quantiles.}
\end{figure}
The resulting quantile fits are shown in Figure \ref{fig:plotErrLev}. It is clear that the bias is positive for $\tau = 0.95$, negative for $\tau = 0.05$ and that its absolute value increases with \code{err}, becoming clearly visible for \code{err} $> 0.1$. However, increasing \code{err} affects the median estimate only slightly. Heuristically, this is because the bias induced by the smoothed loss is greater when the response distribution is more asymmetric around the quantile of interest. 

One might conclude that the loss smoothness should be kept as small as possible, but this leads to computational and numerical stability issues. In fact, decreasing \code{err} tends to slow down computation, and in this example using a very small tolerance increases the computing time dramatically
\begin{Schunk}
\begin{Sinput}
R> system.time(fitBigErr <- qgam(y ~ s(x), data = X, 
+    qu = 0.95, err = 0.05))[[3]]
\end{Sinput}
\begin{Soutput}
Estimating learning rate. Each dot corresponds to a loss evaluation. 
qu = 0.95.............done 
\end{Soutput}
\begin{Soutput}
[1] 0.4
\end{Soutput}
\begin{Sinput}
R> system.time(fitSmallErr <- qgam(y ~ s(x), data = X, 
+    qu = 0.95, err = 0.001))[[3]]
\end{Sinput}
\begin{Soutput}
Estimating learning rate. Each dot corresponds to a loss evaluation. 
qu = 0.95....................done 
\end{Soutput}
\begin{Soutput}
[1] 37.699
\end{Soutput}
\end{Schunk}
More importantly, using a very low value of \code{err} can lead to numerical problems. In fact, the following code
\begin{Schunk}
\begin{Sinput}
R> check(fitSmallErr$calibr, sel = 2)
\end{Sinput}
\end{Schunk}
\begin{figure}[t]
\centering
\begin{Schunk}

\includegraphics[width=\maxwidth]{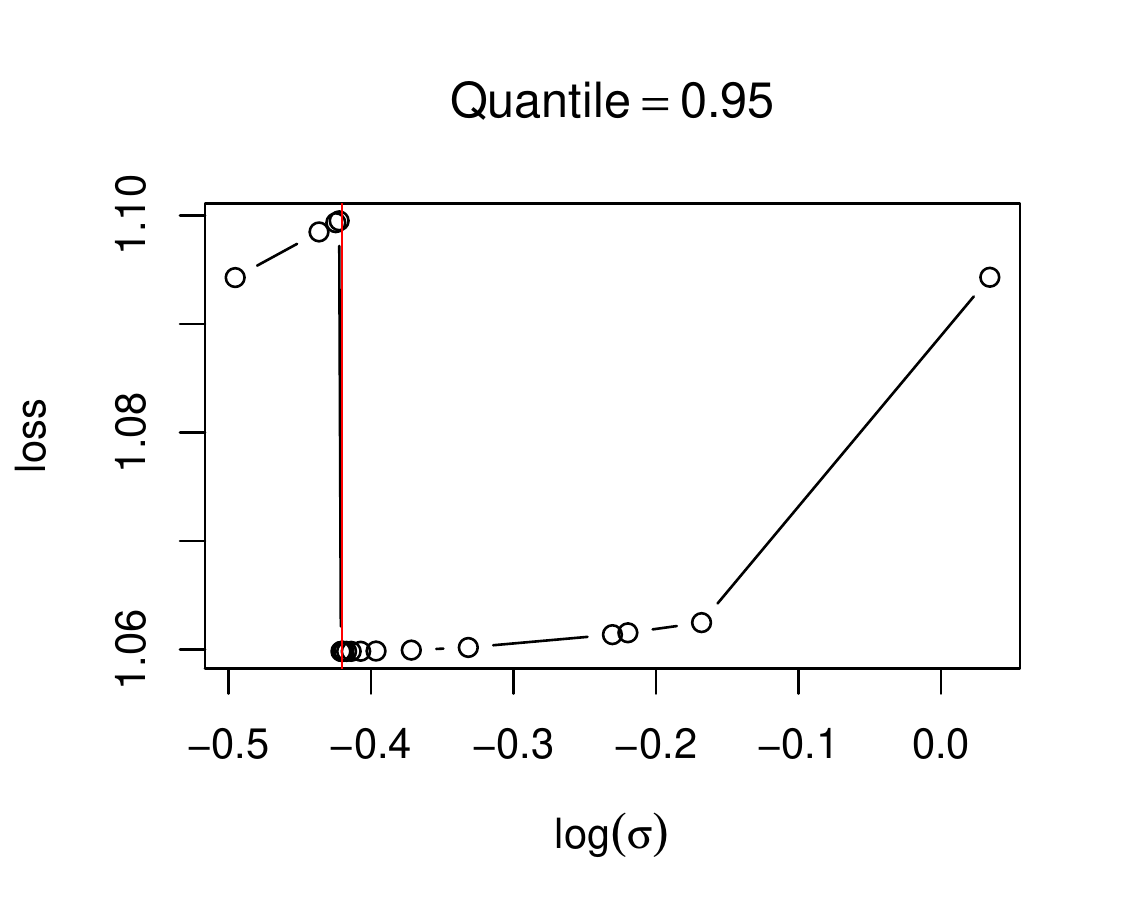} \end{Schunk}
\caption{\label{fig:checkErr001} Plot produced by \code{check.learnFast}, showing the estimated IKL loss at several values of $\log \sigma_0$, when \code{err} = 0.001. The vertical jump in the loss indicates numerical problems.}
\end{figure}
produces the plot in Figure \ref{fig:checkErr001}, which shows how the estimated IKL loss minimized by the outer iteration looks like. The dots indicate the points at which the loss has been evaluated during the outer Brent optimization, and the vertical red line indicates the value of $\log \sigma_0$ which minimizes the loss. Here the loss seems to be discontinuous, which is due to numerical instabilities. In fact, the analogous plot for \code{fitBigErr} (not shown) looks smooth and convex. To produce the diagnostic plot in Figure \ref{fig:checkErr001} we used the generic \code{qgam::check} function, which dispatched to the \code{check.learnFast} method. This is because the output of \code{qgam} contains the results of a call to \code{tuneLearnFast} in the \code{$calibr} slot. The call to \code{check} generates two plots and here we are showing only the second one by choosing \code{sel = 2}.

We can use \code{check} to have also an estimate of the bias attributable to the smoothed loss and information regarding the convergence of the smoothing parameter estimation routine. This is achieved by
\begin{Schunk}
\begin{Sinput}
R> check(fitBigErr)
\end{Sinput}
\begin{Soutput}
Theor. proportion of neg. resid.: 0.95   Actual proportion: 0.952
Integrated absolute bias |F(mu) - F(mu0)| = 0.004319621
Method: REML   Optimizer: outer newton
full convergence after 5 iterations.
Gradient range [2.626207e-05,2.626207e-05]
(score 3187.994 & scale 1).
Hessian positive definite, eigenvalue range [1.874658,1.874658].
Model rank =  10 / 10 

Basis dimension (k) check: if edf is close too k' (maximum possible edf) 
it might be worth increasing k. 

     k'  edf
s(x)  9 5.27
\end{Soutput}
\end{Schunk}
\begin{figure}[t]
\centering
\begin{Schunk}

\includegraphics[width=\maxwidth]{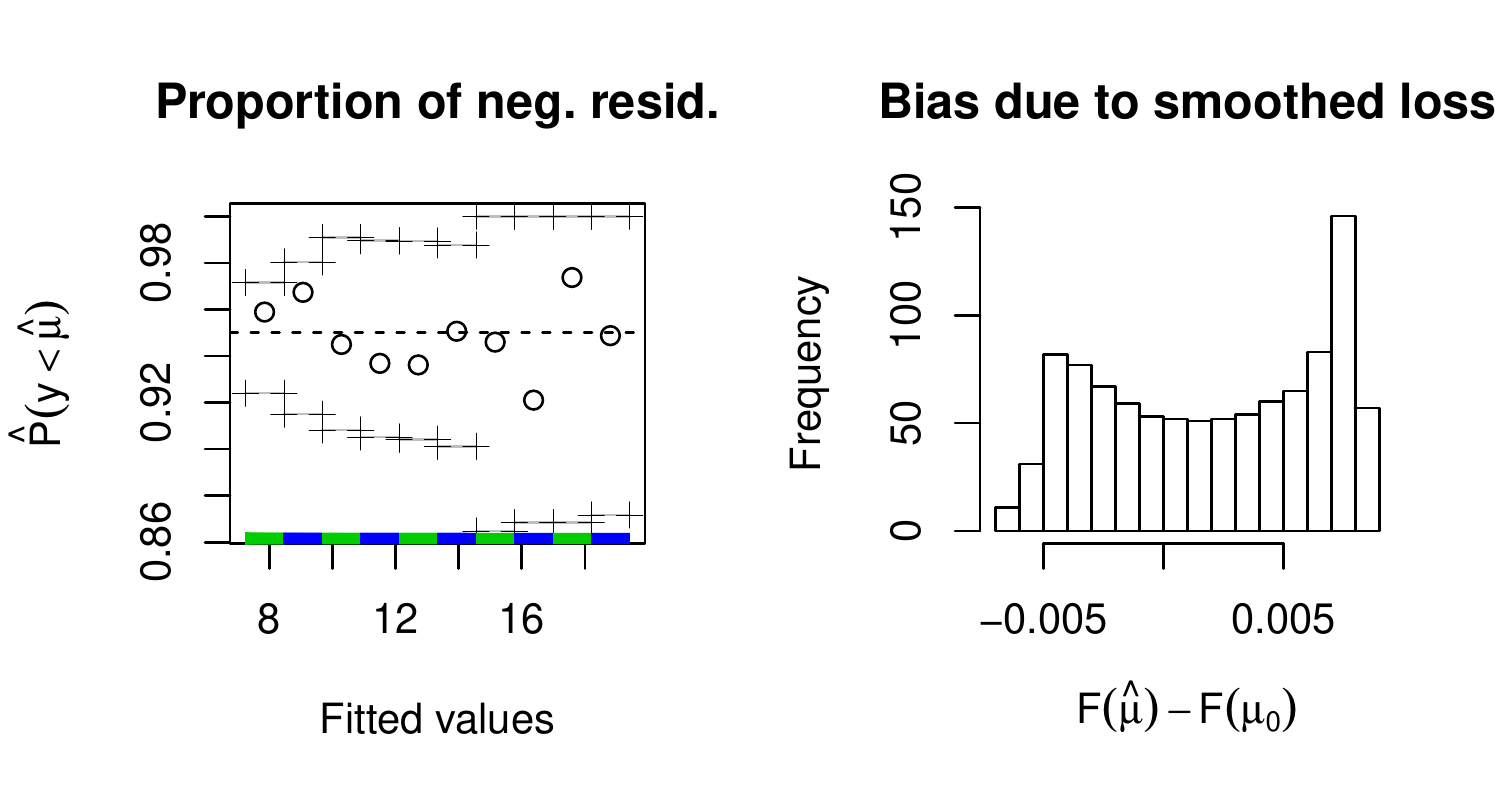} \end{Schunk}
\caption{\label{fig:checkErr001_bis} Diagnostic plots produced by \code{check.qgam}. Left: observed proportion of negative residuals in ten bins, with 95\% reference intervals. Right: histogram of estimated bias attributable to using a smooth loss.}
\end{figure}
which produces the preceding text output and the diagnostic plots shown in Figure \ref{fig:checkErr001_bis}. The text states that the intermediate optimization for selecting $\bm \gamma$ achieved full convergence in few iterations. It reports the range of the entries of the LAML (\ref{eq:LAML}) gradient w.r.t. $\bm \gamma$ at convergence, which we expect to be small, and confirms that the Hessian is positive definite, so that LAML is (at least locally) convex. The text also states that the model has full rank, hence all the regression coefficients $\bm \beta$ are identifiable. The first two lines of the printed text are concerned with the bias due to the use of the smooth ELF loss. The first line says that nearly 95\% of the observations fall below the fitted quantile, which is the percentage we expect for $\tau = 0.95$. The second line reports an estimate of the absolute bias $|F\{\mu^*(\bm x_i)\} - F\{\mu_0(\bm x_i)\}|$, averaged over the observed $\bm x_i$'s. This is much lower than the bound \code{err = 0.05} used to fit this model. Figure \ref{fig:checkErr001_bis} gives more details, with the left plot showing the proportion of negative residuals in a sequence of bins, ordered according to increasing value of $\mu_\tau(\bm x)$. The crosses define 95\% reference intervals for the proportions, obtained using binomial quantiles. The plot on the right is a histogram of all the estimated values of the bias $|F\{\mu^*(\bm x_i)\} - F(\bm x_i)|$, and it shows that the bias is low for all observations. Hence, the text and visual output of \code{check.qgam} confirm that the LAML optimization achieved full convergence and that we should not be too concerned about the bias. The last few lines of the text output indicate that the effective degrees of freedom (\code{edf}) used to model the effect of \code{x} are well below the degrees of freedom available for this smooth effect (\code{k'}). This suggests that the number of basis functions used to model the effect of \code{x} was sufficiently large, while \code{edf} $\approx$ \code{k'} would indicate that the fit is using all the available degrees of freedom, and that we might want to increase the number of basis functions used.  

The results presented in this section and our practical experience on loss smoothness selection and convergence checking with \pkg{qgam} can be summarized in the following set of suggestions:
\begin{itemize}
   \item the automatic procedure for selecting the loss smoothness generally offers a good compromise 
         between statistical bias, variance and numerical stability;
   \item the old default (used in versions of the \pkg{qgam} package lower than \code{1.3.0}) was \code{err = 0.05}, 
         which generally does not lead to unacceptably high levels of bias;
   \item if the calibration loss plotted by \code{check(fittedQGAM$learn)} is irregular, 
         or the text printed by \code{check(fittedQGAM)} does not confirm that \code{full convergence} 
         was achieved, try to increase \code{err};
   \item if you have to increase \code{err} to 0.2 or higher to avoid convergence issues, then
         there might be something wrong with your model (for example, it is missing an important effect);       
   \item the bias attributable to the adoption of a smoothed ELF loss can be estimated using \code{check(fittedQGAM)};
   \item you might get messages saying that \code{outer Newton did not converge fully} during estimation. These are generated 
         by \code{mgcv::gam} during LAML maximization, and should not be problematic as long as the calibration loss 
         is smooth (which you can check using \code{check(fittedQGAM$calibr)}) and \code{check(fittedQGAM)} 
         states that \code{full convergence} was achieved;
   \item in preliminary studies, that is when you are exploring different model structures and you are not yet interested 
         in getting the most accurate estimates, do not decrease \code{err} too much as it considerably slows down computation;
   \item setting \code{err} too low is generally not a good idea. In fact, \code{err} is an approximate upper bound
         on the bias, the latter being generally much lower than this parameter suggests, and it is arguably better to have a small 
         amount of bias than numerical problems. As stated above, the default loss smoothness selection procedure, used by default
         in \code{qgam} and \code{mqgam}, typically provides stable fits and a near optimal bias-variance tradeoff.
\end{itemize}

\newpage

\subsection{Derivation of the asymptotic bias} \label{app:approxQualProof}

To simplify the notation, indicate $\tilde{p}_\tau(y - \mu)$ with $\tilde{p}_\tau(y)$. We start from
\begin{equation*}
F(\mu^*) - F(\mu_0) = \int \mathbbm{1}(y\leq\mu^*)f(y)dy - \tau = \int \bigg \{ \frac{\partial \log\tilde{p}_\tau(y)}{\partial \mu}\bigg|_{\mu = \mu^*} - \frac{\partial \rho_\tau(y)}{\partial \mu}\bigg|_{\mu = \mu^*} \bigg \} f(y)\, dy,
\end{equation*}
where $f(y)$ is the p.d.f. of $y$ and we used the fact that $\int \partial \log\tilde{p}_\tau(y)/\partial \mu|_{\mu = \mu^*}f(y)dy = 0$, by definition of $\mu^*$. 
We proceed to bound the right hand side (r.h.s.) from above. For any $\mu$, simple manipulations lead to
\begin{equation} \label{eq:symmetric}
\int \bigg \{ \frac{\partial \log\tilde{p}_\tau(y)}{\partial \mu} - \frac{\partial \rho_\tau(y)}{\partial \mu} \bigg \} f(y)\, dy = \int \bigg \{ \Phi(y|\mu, \lambda \sigma) - \mathbbm{1}(y>\mu) \bigg \} f(y)\, dy,
\end{equation}
where $\mathbbm{1}(\cdot)$ is the indicator function and $\Phi(y|\mu, \lambda \sigma)$ is the c.d.f. of a logistic r.v. with mean $\mu$ and scale $\lambda\sigma$. Then we have
\begin{eqnarray*}
|F(\mu^*) - F(\mu_0)| &\leq&  \int \bigg | \Phi(y|\mu^*, \lambda \sigma) - \mathbbm{1}(y>\mu^*) \bigg | \sup_y{f(y)} \, dy \\
&=& 2 \sup_y{f(y)} \int_{-\infty}^{\mu^*} \Phi(y|\mu^*, \lambda \sigma)\, dy,
\end{eqnarray*}
where the second equality holds due to the symmetry of the integrand around $\mu^*$. Finally, the substitution $z = (y-\mu^*)/\lambda\sigma$ leads to
\begin{eqnarray*}
|F(\mu^*) - F(\mu_0)| &\leq& 2 \lambda \sigma \sup_y{f(y)} \int_{-\infty}^{0} \frac{1}{1+e^{-z}}\, dz \\ 
&=& 2\log{(2)}\lambda\sigma \sup_y{f(y)}.
\end{eqnarray*}
Note that the r.h.s. of (\ref{eq:symmetric}) makes it clear that, if $f(y)$ is symmetric around $\mu^*$, then $|F(\mu^*) - F(\mu_0)| = 0$ and there is no bias.

\end{appendix}

\end{document}